\documentclass[traditabstract,oldversion]{aa} 

\usepackage{graphicx}

\usepackage{txfonts}
\def\bibl{\parindent=0pt \hangindent=0.150 in }
\begin{document}

\title{ Search for  Cold Debris Disks around  M-dwarfs.  II}

\author{J.-F. Lestrade 
       \inst{1}
       \and 
        M. C. Wyatt \inst{2}
       \and
       F. Bertoldi\inst{3}
       \and
        K. M. Menten\inst{4}
       \and
        G. Labaigt\inst{5}
}

\offprints{J.-F. Lestrade, e-mail : jean-francois.lestrade@obspm.fr}

\institute{Observatoire de Paris - CNRS, 77 av. Denfert Rochereau, F75014, Paris, France\\
\email{jean-francois.lestrade@obspm.fr}
\and 
Institute of Astronomy, University of Cambridge, Cambridge, CB3 OHA, UK\\
\email{wyatt@ast.cam.ac.uk}
\and
Argelander Institute for Astronomy, University of Bonn, Auf dem H\"ugel 71, Bonn, D-53121, Germany\\
\email{bertoldi@astro.uni-bonn.de}
\and
Max-Planck-Institut f\"ur Radioastronomie, Auf dem H\"ugel 69, Bonn D-53121, Germany\\
\email{kmenten@mpifr-bonn.mpg.de}
\and
Ecole Normale Sup\'erieure de Cachan, 61 avenue du Pr\'esident Wilson, F94235 - Cachan - France\\
\email{glabaigt@rip.ens-cachan.fr}  
}

\date{Received 22nd April 2009 ; accepted 6th July 2009}

\abstract{
Although 70\% of the stars in the Galaxy are M-dwarfs, thermal emission searches for cold debris disks have been 
conducted mostly for A-type and solar-type stars.  
We report on new $\lambda=1.2$~mm continuum observations of thirty M-dwarfs, using the MAMBO-2 bolometer array camera 
at the IRAM 30m telescope.
For a statistical analysis, we combine these data with our prior SCUBA and MAMBO-2 observations of 20 other M-dwarfs. 
Our total sample divides in M-dwarfs in moving groups, with relatively young ages, and in nearby M-dwarfs with unknown ages. 
Only one cold debris disk  (GJ842.2) was detected significantly.
We compare the implied disk abundance constraints with those found in two comparable submillimeter surveys 
of 10 to 190 Myr old A- and FGK-type stars.
For the 19 youngest (ages less than 200 Myr) M-dwarfs in our sample, we derive a disk fraction of  $5.3^{+10.5}_{-5.0}~\%$,
compared to  $15 ^{+11.5}_{-11.5}~\%$ for FGK-stars and $22^{+33}_{-20}~\%$ for A-stars. 
Hence, for this age group, there is an apparent trend of fewer cold disks for later stellar 
types, {\it i.e.}, lower star masses. Although its statistical significance is marginal, this trend is strengthened 
by the  deeper sensitivity of observations in the M-dwarf sample. We derive a cold disk fraction of $< 10~\%$ for the older (likely a few Gyr)
 M-dwarfs in our sample. Finally, although inconclusively related to 
a debris disk, we present the complex millimeter structure found around  the position of the M1.5 dwarf GJ526 in our sample.  

}

\titlerunning{ Debris Disks around  M-dwarfs}
\authorrunning{Lestrade et al. }

\maketitle

\section{Introduction}
\label{Introduction}

Cold debris disks around main sequence stars are left-over planetesimals (comets) that could not agglomerate into larger
planets  during the initial phase of planet formation. 
They are assembled as a belt in the periphery of the system in a manner analogous to the Kuiper Belt. 
The study of debris disks, including warm disks such as the asteroid belt in our Solar System, 
advances our knowledge of the origin and evolution of planetary systems around other stars, similarly to 
the study of the  Kuiper Belt.  It is well recognised that its
present-day structure and dynamics retain important information on the formation and evolution 
of the Solar System. For example, the low mass and expanded size of the present-day Kuiper Belt can be traced back to the outward migrations
of the giant planets, which exchanged orbital energy with  an initially more compact
and more massive disk  (Hahn \& Malhotra 1999, Tsiganis et al. 2005, Morbidelli et al., 2005, and Gomes et al. 2005).

Our current understanding of debris disks has been recently 
reviewed by Wyatt (2008). Mutual collisions between planetesimals in debris disks produce second-generation dust grains that are 
observable through  their thermal emission or scattered light.
Since the discovery of a first debris disk around the A0 main sequence star Vega by IRAS (Aumann et al. 1984), debris disks 
have been searched for photometrically with the infrared (IR) satellites IRAS, ISO and Spitzer. In such observations,   
any flux excess above the photospheric level is interpreted as emission from warm  ($50-100$~K) circumstellar dust.

Through Spitzer observations it was found that  $ 32 \pm 5 \%$  of 160 A-dwarfs show a 70~$\mu$m excess (Su et al. 2006), while
only $16^{+2.8}_{-2.9}\%$  of 225 observed FG-dwarfs show excess emission (Bryden et al. 2006; Trilling et al. 2008). 
Submillimeter photometry has shown that some A-to-G type stars with no detectable IRAS excess do however show cold ($10 - 50$~K) dust 
emission  (Wyatt et al 2003; Najita \& Williams 2005).
Imaging of scattered light with the  HST 
and  of thermal continuum emission with SCUBA has measured disk radii between 50 and 150~AU  
(Smith \& Terrile 1984; Kalas, Graham \& Clampin 2005; Holland et al. 1998; Greaves et al. 2005) ; 
22 debris disks have been imaged presently \footnote{http://astro.berkeley.edu/~kalas/disksite/pages/gallery.html}.  
Azimuthal structures have been detected in a few of these disks and are thought to be caused by dust associated with planetesimals trapped in mean
motion resonance with an orbiting planet  (Wyatt 2003, 2006, Reche et al. 2008),  or by dynamical perturbations from a distant 
stellar companion or passing stars (HD141569A: Augereau \& Paploizou 2004).  

Although low-mass M-dwarfs are the most populous (70\%) stars in the Galaxy, they have so far received little attention, mostly because their low luminosity
makes the thermal emission and scattered light from their disks more difficult to detect. 
In a SCUBA survey of young stars of the  {$\beta$~Pic 
moving group} and of the Local Association, Liu et al. (2004) detected
the first two debris disks around M-dwarfs, AU~Mic and GJ182.
Using SCUBA and MAMBO-2, Lestrade et al. (2006) surveyed  32 relatively young M-dwarfs of moving groups
and newly detected one disk around the M0.5 dwarf GJ842.2. Using Spitzer, Gautier et al. (2007)  
surveyed 62 nearby M-dwarfs at 24~$\mu$m, and subsamples of 41 at  70~$\mu$m and of 20 at 160~$\mu$m, and found no firm detection. 
AU~Mic was also imaged in scattered light, revealing an edge-on, structured disk (Liu, 2004 and Krist et al., 2005).    

Here we present new MAMBO-2 observations of nearby M-dwarfs, and combined them with our previous survey
to analyze a total sample of 50 M-dwarfs (Table~1) in terms of their cold debris disks abundance.
 
We present the new M-dwarfs surveyed in \S2, we describe the observations in \S3, and results for debris disks but also for  
background sources in the fields of some of the M-dwarfs in \S4. Finally, we compare the detected fraction of 
cold debris disks around the M-dwarfs
of our sample with the fractions for other stellar types to discuss whether or not the characteristics  of M-dwarfs
$-$ mass and luminosity lower than solar $-$ impact the formation of debris disks around them or their detectability.

\section{Sample of newly observed M-type dwarfs}

To complement our first survey of M-dwarfs, which are in moving groups of ages $<$ 600 Myr (Lestrade et al 2006), we
observed the most nearby M-dwarfs, irrespective of age.  We selected single M-dwarfs at a distance less
than 6~pc and at $\delta > -11^{\circ}$, and added five M-dwarfs binaries and  six single M-dwarfs between 6 and 10 pc
that are in common with the Spitzer survey by Gautier et al. (2007).
The five binaries in our sample are: GJ725 (M3 and M3.5) separated by $15''$~$(53~AU)$,
GJ234 (M4.5 and M8), separated by $1''$~$(5~AU)$, GJ412 (M2 and M6), separated by $32''$~$(160~AU)$, 
GJ569 (M2.5 and M8.5), separated by $1''$~$(10~AU)$, and GJ65 (M5.5  and M5.5), separated by $2''$~$(5.2~AU)$. 
These angular separations are so small that a single MAMBO-2 map can cover both components. 
The ages of the targeted near-by M-dwarfs are presently unknown.  

\section{Observations}

The  diameter  usually adopted for debris disks is 120~AU, which for near-by stars is larger than the IRAM 30-meter telescope beam of $10.7''$ FWHM
at $\lambda=1.2$~mm. We have therefore imaged each field with the
117-channel Max-Planck Bolometer array (MAMBO-2; Kreysa et al. 1998) of the 30m telescope on Pico Veleta, 
Spain (2900~m). MAMBO-2 has a half-power spectral bandwidth from 210 to
290 GHz, with an effective frequency  centered at 250~GHz (1.20~mm) for thermal emission spectra. 
The effective FWHM beam  is $10.7''$, and the undersampled field of view of the array is $4'$. 
We used the standard on-the-fly mapping technique, where one map is made
of 41 azimuthal subscans of 60 sec each, with a scanning velocity of $\rm 4''/sec$ and  an elevation incremental step of $4''$, 
while chopping the secondary mirror at 2 Hz by $60''$ in azimuth.  The bolometers are arranged in a hexagonal pattern 
with a beam separation of $22''$. This scanning pattern produces time streams of data that are converted 
to a fully sampled spatial map with $3.5''$ pixels. Our observations
were done within pooled observing runs spread over the winter and summer periods  
from 2005 to 2007. Atmospheric conditions were generally good during the observations,
with typical zenith opacities between 0.1 and 0.3 at 250 GHz and low sky noise. 
The telescope pointing was checked before and after each map by using the same nearest bright point source, and
was found to be stable to better than  $3''$, except in a few occasions for which we discarded the data. 
The absolute flux calibration is based on observations of several standard calibration sources, including
planets, and on a tipping curve (sky dip) measurement of the atmospheric opacity once every few hours. 
The resulting absolute flux calibration uncertainty is estimated to be about 10\% (rms).

The data were analyzed using the mopsic software package written by R. Zylka at IRAM.
The double-beam maps were combined to a single map using the shift-and-add procedure. Compared to a proper image 
restoration this produces maps with about a factor 2 better sensitivity, at the expense of no sensitivity to emission 
structures in scan direction that are larger that the wobbler throw of $60''$.
In our sample, the shortest integration time  per field
is 30 minutes, yielding an rms noise level of $\rm \sim 2.5~mJy/11''$~beam in the central part
of the map ($r < 60''$), steadily rising  to $\rm \sim 5~mJy/11''~beam$ at its edge ($r \sim 140''$). 
This non-uniform noise across the maps results from the fact that the  scanned field is about twice
as large as the bolometer array size, so that more data is taken in the central part of the map
than near the edges. Due to this non-uniformity, it is judicious to present Signal-to-Noise maps
rather than intensity maps.  The longest observation, by far, had  20 hour  duration and was on  GJ628, 
yielding an  rms noise level of $\rm \sim 0.5~mJy/11''~beam$
in the central part of the map and   $\rm \sim 1.2~mJy/11''~beam$ at  $r \sim 140''$.
The sky area covered by each  map corresponds to $\sim$ 1700 independent antenna beams. The residual noise in the map was found to be nearly Gaussian.
Hence, a $4\sigma$ detection in our maps is statistically significant.
Below, we describe the procedure we used to extract discrete sources and to search for extended emission.

\begin{table*}
\begin{center}
\caption{The 50 M-dwarfs of our two (sub)mm surveys. Survey I was already presented by Lestrade et al. (2006) but its data were 
 included in the statistical analysis of this paper. New data were acquired in survey II with MAMBO-2.
Selection criteria are  ages $<$ 600 Myr for survey~I and the nearest M-dwarfs  but of undetermined ages for survey~II. Ages are based on 
Moving Groups (Local Association ($20 - 150$~Myr), IC2391 ($35 - 55$~Myr), AB Dor (100-125 Myr), Castor (200 Myr), Ursa Maj (500 Myr)  and Hyades (600 Myr)) 
identified by Montes et al. (2001) 
and Zuckerman \& Inseok Song (2004a,b). Some stars were observed in both surveys. }
\begin{tabular}{l|lcccll} \hline\hline\\\
     Name   & Spectral   & Binarity   &    Selection         &   Observed in      &   Bolometer  &  Publication         \\
            &            &            &    criterium         &   survey \#        &   arrays     &                      \\\\\hline\\\
 GJ82       &   M4       &  Single    &    $20 - 150$~Myr    &    I               &  SCUBA       &  Lestrade et al. (2006) \\
 GJ212      &   M0.5     &    ''      &       ''             &    I               & ~~~~  ''      &        ''             \\  
 GJ507.1    &   M1.5     &    ''      &       ''             &    I               & ~~~~   ''        &        ''             \\
 GJ696      &   M0       &    ''      &       ''             &    I               & ~~~~   ''        &        ''             \\  
 GJ876      &  M4        &    ''      &       ''             &    I               & MAMBO       &        ''             \\  
 GJ628      &  M3.5      &    ''      &       ''             &    I               & ~~~~   ''        &        ''             \\  
 GJ402      &  M4        &    ''      &       ''             &    I               & ~~~~   ''        &        ''             \\  
 GJ234      &   M4.5+M8  &  Binary    &       ''             &    I \&  II        & ~~~~  ''        &     Lestrade et al. (2006) and this work      \\
 GJ285      &   M4.5     &  Single    &       ''             &    I               &  SCUBA \& MAMBO & Lestrade et al. (2006)  \\  
 GJ393      &   M2       &    ''      &       ''             &    I               & ~~~~    ''        &        ''             \\
 GJ9809     &   M0       &    ''      &       ''             &    I               &  ~~~~   ''        &        ''             \\
 GJ875.1    &   M3       &    ''      &    $35 - 55$~Myr     &    I               &  SCUBA       &        ''              \\
 GJ856A     &  M3        &    ''      &    100-125 Myr       &    I               &  MAMBO       &         ''             \\    
 GJ277B     &   M3.5     &    ''      &     200 Myr          &    I               & SCUBA       &        ''              \\
 GJ842.2    &   M0.5     &    ''      &      ''              &    I               & ~~~~   ''        &        ''    \\
 GJ890      &   M2       &    ''      &      ''              &    I               & ~~~~    ''       &        ''               \\ 
 GJ1111     &   M6.5     &    ''      &      ''              & I \&  II           &  MAMBO      & Lestrade et al. (2006) and this work \\
 GJ408      &   M2.5     &    ''      &      ''              & I \&  II           & ~~~~   ''        &         ''             \\ 
 GJ4247     &   M4       &    ''      &      ''              & I \&  II           &  SCUBA \& MAMBO &      ''             \\  
 GJ447      &   M4.0     &    ''      &     500 Myr          &  I \&  II          &  MAMBO          &        ''            \\
 GJ625      &   M1.5     &    ''      &      ''              &  I \&  II          & ~~~~  ''         &         ''           \\ 
 GJ569      & M2.5+M8.5  &  Binary    &      ''              &  I \&  II          & ~~~~  ''         &         ''           \\
 GJ873      &  M3.5      &  Single    &      ''              &    I               & ~~~~  ''         &     Lestrade et al. (2006)       \\ 
 GJ65       &  M5.5+M5.5 &  Binary    &     600 Myr          &    I               & ~~~~  ''         &        ''     \\  
 GJ3379     &  M4        &  Single    &      ''              &    I               & ~~~~    ''       &        ''             \\      
 GJ849      &  M3.5      &    ''      &      ''              &    I               & ~~~~   ''        &        ''            \\   
 GJ791.2    &  M4.5      &    ''      &      ''              &    I               & ~~~~   ''        &        ''            \\ 
 GJ109      &   M3.5     &    ''      &      ''              &    I  \& II        & ~~~~    ''        &    Lestrade et al. (2006) and this work   \\\\\hline\\
 GJ699      &   M4.0     &  Single    &      1.82 pc         &    II              &   MAMBO     &      this work     \\  
 GJ406      &   M6.0     &    ''      &      2.38 pc         &    II              & ~~~~   ''        &         ''           \\            
 GJ411      &   M2.0     &    ''      &      2.54 pc         &    II              & ~~~~   ''         &         ''           \\
 GJ905      &   M5.5     &    ''      &      3.16 pc         &    II              & ~~~~   ''        &         ''           \\
 GJ725      &   M3+M3.5  &  Binary    &      3.57 pc         &    II              &  ~~~~  ''        &         ''           \\
 GJ54.1     &   M4.5     &  Single    &      3.72 pc         &    II              &  ~~~~  ''        &         ''           \\
 GJ273      &   M3.5     &    ''      &      3.79 pc         &    II              &  ~~~~  ''        &         ''           \\
 GJ83.1     &   M4.5     &    ''      &      4.44 pc         &    II              &  ~~~~   ''       &         ''           \\
 GJ687      &   M3.0     &    ''      &      4.53 pc         &    II              &  ~~~~   ''       &         ''           \\
 LHS292     &   M6.5     &    ''      &      4.54 pc         &    II              & ~~~~    ''       &         ''           \\
 GJ1002     &   M5.5     &    ''      &      4.69 pc         &    II              & ~~~~    ''       &         ''           \\
 GJ412      &   M2+M6    &  Binary    &      4.83 pc         &    II              & ~~~~   ''        &         ''           \\
 GJ388      &   M3.0     &  Single    &      4.89 pc         &    II              & ~~~~   ''        &         ''           \\
 GJ445      &   M3.5     &    ''      &      5.38 pc         &    II              & ~~~~   ''        &         ''           \\
 LHS1723    &   M4.5     &    ''      &      5.43 pc         &    II              & ~~~~   ''        &         ''           \\
 GJ526      &   M1.5     &    ''      &      5.43 pc         &    II              & ~~~~   ''        &         ''           \\
 GJ251      &   M3.0     &    ''      &      5.57 pc         &    II              & ~~~~   ''        &         ''           \\ 
 GJ205      &   M1.5     &    ''      &      5.71 pc         &    II              &  ~~~~  ''        &         ''           \\
 GJ213      &   M4.0     &    ''      &      5.87 pc         &    II              & ~~~~   ''        &         ''           \\  
 GJ908      &   M1.0     &    ''      &      5.93 pc         &    II              & ~~~~    ''       &         ''           \\
 GJ581      &   M3       &    ''      &      6.27 pc         &    II              &  ~~~~   ''       &         ''           \\
 GJ102      &   M4       &    ''      &      7.75 pc         &    II              & ~~~~    ''       &         ''           \\
\\\hline   
\end{tabular}
\end{center}
\end{table*}

\section{Results}

Although initially intended, our complete survey of 50 M-dwarfs turns out not to be flux-limited, which makes
its statistical interpretation not straightforward.
First we present the discrete millimeter sources detected in four MAMBO-2 maps (Figs 1, 2 and 3, and Table~2), and we discuss
the nature of the intriguing cluster of sources around GJ526. Second, we present the  
deep search for faint debris disks made by averaging intensities over an effective area in the 42 MAMBO-2 maps 
of our complete survey (Tables~3 and 4). The 8 other M-dwarfs were observed in wide photometry with SCUBA (Table~5)
and were already discussed by Lestrade et al. (2006).  
We  use the complete sample of 50 M-dwarfs to estimate the fraction of cold debris disks around M-dwarfs and upper limits 
of their fractional dust luminosities. Three stars (GJ285, GJ393 and GJ4247) are in common between Tables~3, 4 and 5.

\subsection{Discrete sources and structure around GJ526.}

\begin{table*}
\caption{Sources found in our MAMBO-2 survey ($S/N > 4$). The IRAM30m beam is $11''$ at $\lambda=1.2$~mm. 
The flux density of MM184222+593828 is provided as lower and upper limits because the source is located close 
to  one of  the borders of the map where some systematics remain in the data  (see text). }
\begin{tabular}{l|l|ccccc} 
\hline\hline\\\
 Source name     &  Star Field & $\alpha$(J2000) $^a$ & $\delta$(J2000) $^a$ &    Integrated flux density $^b$   & $S/N$ &  Source  \\
                 &             &                      &                      &     at 1.2~mm (mJy)  &       & FWHM  \\\\
\hline\\
MM145428+160439  & GJ569       &  14 ~54 ~28.2        &  16 ~04 ~39          &    $ 5.3 \pm 1.1$    & 4.9 &  $11''$    \\
                 &             &                      &                      &                      &     &            \\
MM184222+593828  & GJ725       &  18 ~42 ~22.7        &  59 ~38 ~28          &     $30 - 63$          & $6.7 - 7.3$ &  $13'' - 15''$     \\
MM184253+593756  & GJ725       &  18 ~42 ~53.3        &  59 ~37 ~56          &    $ 9.7 \pm 1.9 $   & 5.1 &  $11''$    \\    
                 &             &                      &                      &                      &     &            \\ 
MM134539+145139  &  GJ526      &  13 ~45 ~39.0        &  14 ~51 ~39          &    $3.2 \pm 0.7 $    & 4.4 &  $11''$    \\
MM134540+145446  &  GJ526      &  13 ~45 ~40.7        &  14 ~54 ~46          &    $4.3 \pm 0.8 $    & 5.2 &  $11''$    \\
MM134541+145417  &  GJ526      &  13 ~45 ~41.3        &  14 ~54 ~17          &    $6.3 \pm 1.0 $    & 6.1 &  $11''$    \\
MM134543+145317  &  GJ526      &  13 ~45 ~43.1        &  14 ~53 ~17          &    $3.0 \pm 0.7 $    & 4.9 &  $11''$     \\
MM134546+145240  &  GJ526      &  13 ~45 ~46.7        &  14 ~52 ~40          &    $5.6 \pm 0.7 $    & 8.6 &  $11''$    \\
                 &             &                      &                      &                      &     &            \\ 
MM163007-123942  &  GJ628 $^c$ &  16 ~30 ~07.6        & $-$12 ~39 ~42        &    $7.3 \pm 1.2 $    & 6.2 &  $14\pm 4''$    \\
MM163022-123925  &  GJ628 $^d$ &  16 ~30 ~22.3        & $-$12 ~39 ~25        &    $3.8 \pm 0.9 $    & 4.4 &  $17\pm 5''$    \\
MM163019-123830  &  GJ628 $^e$ &  16 ~30 ~19.7        & $-$12 ~38 ~30        &    $4.1 \pm 0.8 $    & 4.9 &  $11''$      \\
MM163015-123911  &  GJ628 $^f$ &  16 ~30 ~15.6        & $-$12 ~39 ~11        &    $4.8 \pm 0.8 $    & 6.2 &  $11''$     \\
MM163013-124057  &  GJ628 $^g$ &  16 ~30 ~13.7        & $-$12 ~40 ~57        &    $4.3 \pm 1.0 $    & 4.8 &  $11''$     \\\\
\hline
\end{tabular}

\smallskip

\scriptsize

 $^a$ The MAMBO coordinate uncertainties are $\sim 3.5''$.

 $^b$ Flux density (mJy) is integrated under the fitted 2D-Gaussian. 

 $^c$, $^d$, $^e$, $^f$, $^g$ : these sources were found in survey I ;  ($^c$)  is GJ628-W,  ($^d$) is GJ628-E,   ($^e$) is GJ628-NE,  ($^f$) is GJ628-NW, 
and ($^g$) is GJ628-SW in Fig~2 and Table~3  of Lestrade et al. (2006). Note that there are typographic errors in 
 the $minute$ column of declinations in Table~3 of Lestrade et al. (2006). The positions of the GJ628 sources given  
now are correct and supersede this first publication. Also, the flux densities in this first publication were
peak flux densities  while they are integrated flux densities now.

\end{table*}

\subsubsection {Source extraction}

In Table~2, we summarize the characteristics of the discrete  sources detected with $S/N > 4$ in the 
42 fields mapped with MAMBO-2, each being $\sim 400'' \times 400''$ in size and centered on the position of 
an M-dwarf. The source extraction was done by searching each map for any pixel with $S/N > 4$ to select 
a block of $7\times7$ pixels centered on it. Then, we  minimized  $\chi^2$ 
between the 2-D~Gaussian $ F \times {\rm exp}\big(-0.5 \times \big[{(x-x_0)}^2 + {(y-y_0)}^2~\big] / {\sigma}^2\big)$  
and the measured intensities over this block by varying the peak flux $F$, the parameter $\sigma$ (FWHM$=2.35 \sigma$) 
and the peak position $(x_0, y_0)$ by less than a pixel from the $S/N > 4$ pixel.  
We used the mean brightness computed from the  32 pixels surrounding  the $7 \times 7$ block to remove
any constant level.  With this scheme, this level is determined from pixels that are far enough 
from the 2-D~Gaussian peak to be at the map floor since the Gaussian  FWHM is 3 to 5 pixels 
($11''- 17''$) for our sources.
This correction  amounted to between 5\% and 20\% of the integrated flux density.
The integrated flux densities  of sources in Table~2 are $S_{int}=2 \pi \sigma F$.
We have extracted 13 discrete sources with $S/N~>~4$ for $F$ in 4 fields out of the 42  MAMBO-2 maps~; 
8 sources are newly found in the  fields of  GJ526,  GJ725 and  GJ569.  They are shown in Fig.~1,~2, and 3.
The 5 others sources are in the field of  GJ628 as already  reported in Lestrade et al. (2006)~;
 typographic errors in their coordinates in this first publication are corrected in Table~2.      
In this Table, the lowest integrated flux density is $3.0\pm0.7$ mJy and the highest is between 30 and  63~mJy.  There is no 
optical  counterpart within $7''$ ($2\sigma$ MAMBO-2 position error) to any of these  sources in the  USNO-B1 catalogue
(Monet et al. 2003), except for MM163007-123942 (GJ628W) at the $1 \sigma$ level in position (object I mag = 18.92) 
as already reported in Lestrade et al. (2006). Additionally, we have searched the fields of GJ526 and 
GJ569 that have been imaged in the Sloan Digital Sky Survey (SDSS). In the field of GJ526,  there is
only one SDSS object (z mag = 20.4) which is  as close as $7''$ from MM134540+145446.
However,  the confusion limit is reached over such a  separation
since the density  of SDSS objets  is $\sim$22000
objects/deg$^2$ as measured over the $4' \times 4'$ area centered on the star position.   
In the field of GJ569 ($\sim$33000 SDSS objects/deg$^2$), there are two SDSS objets (mag u =19.9, g=21.6) 
as close as $8''$ from MM145428+160439, but again confusion prevents identification with the MAMBO-2 millimeter source. 
We have found no radio counterpart within the $3\sigma$ MAMBO-2 position uncertainty 
($10.5''$) in the NVSS catalogue (Condon et al., 1998) to the
flux density limit of 2.5 mJy at 1.4~GHz, except for MM163007$-$123942 (GJ628W) at the $1.7 \sigma$ level in position 
(object flux density = $ 5.6\pm 0.5$ mJy) as already reported
in Lestrade et al. (2006). Finally, we have found no IR counterpart within the $3\sigma$ MAMBO-2 position uncertainty 
in the  2MASS All-Sky Catalog of Point Sources   (Skrutskie et al., 2006).

\subsubsection{A bright millimeter source}

The source MM184222+593828 in the field of GJ725  has a  flux density between 30 and 63 mJy 
(Table~2), which is bright compared with millimeter sources found previously in empty fields mapped with MAMBO-2
(Greve et al., 2004, Bertoldi et al., 2007 and Voss et al. 2006). These lower and upper limits were obtained
by using different atmospheric correlation lengths  in the skynoise reduction
within the {\sl mopsic} data reduction package, yielding  similar detections of  $S/N \sim 7$. Systematics 
affects the accuracy of the flux measurement for this source, which is near the edge of the map.   
This  source is  possibly resolved with  a source FWHM between $12''$ and  $15''$ 
depending on the reduction parameter used. Although this is rare, it has recently been shown that submm galaxies can have
multiple components (Tacconi et al., 2008).  
We shall discuss this source together with complementary  observations in a forthcoming paper.

\subsubsection{Structure around GJ526}

The map of GJ526 is shown in Fig.~1. Five sources are detected with $S/N~>4$ and an ellipse almost centered 
on the star  can connect all but the one to the SW. 
Precisely, there are four robust sources with $S/N >4$ that can be connected, 
and  an additional one, farthest to the SE that is only detected at  3.5 times the rms noise level,  and possibly 
real since statistically less than 1 positive noise peak is expected at this level over the whole map if dominated by gaussian noise.
To estimate whether this is the case or not, we have identified the negative noise spikes $S/N < -3$ and  the positive spikes 
$3 < S/N < 4$ in our $440'' \times 420''$ map ($\sim$ 2050 beams). The total of 7 such noise spikes found in the map is consistent 
with the  6  theoretically expected for $|S/N| > 3$ for  gaussian noise. 
Also, the presence of the SE source is found robust  since it appears at the same position in the two independent maps 
we made by halving the data. There is also a $\sim 3 \sigma$ source to the West close to the ellipse but this 
source is not robust to data selection. So we have disregarded it.   

The ellipse is offset by two pixels in right-ascension from the position of the star, 
its semi-major and semi-minor axes are  $a=98''$ 
and $b=15''$, and its orientation is PA=$-30^{\circ}$. This structure might be the projected ring 
of a clumpy  debris disk, its inclination being  $81^{\circ}$ from the  plane of the sky. 
The  offset between the star and the center of the ellipse   is only 2 pixels  and might be caused by 
position uncertainties or/and by real source structure.
The 5 connected sources appear not to be embedded in any extended emission  as expected for 
a debris disk though. The  mean brightness over a rectangular box  $220'' \times 40 ''$ (92 beams)
oriented at PA =$-30^{\circ}$ and centered on the star  is $0.21 \pm 0.073$ mJy/$11''$~beam, 
{\it i.e.}  $\sim 3\sigma$. Actually, this mean brightness matches the mean of the 5 flux densities of the connected sources within
the box, indicating that any extended emission must have a brightness  $\rm < 3 \times 0.073 mJy/11''$~beam at 1.2~mm.

We now test whether or not such a source cluster can arise from the distribution of background sources.
We carry out a first test to estimate  the probability that  the connected sources  around GJ526 can cluster 
as tightly as they do  in Fig~1 if they were background sources. 
For this test, we use the  statistical analysis of spatial point patterns developed  by Diggle (2003). 
This analysis is based on the nearest neighbour distance, defined as the distance between a point 
(a source for us) and its  nearest neighbour. For GJ526, all the nearest neighbour distances for the 5 sources 
connected by the ellipse are $ < 60''$.   
The probability that $k$ distances $x$ be $ < 60''$ among $N$ distances if the mean frequence of occurence 
for $x < 60''$ over the whole sky is $f$,  is given by the binomial distribution  $ B_k  = \big(^N_k \big)~f^k~(1-f)^{N-k}$.
We can derive the mean frequence of occurence  $f$  from  the three    
empty fields (ELAIS N2, Lockman Hole and COSMOS) mapped by MAMBO-2 (Greves et al., 2004 and  Bertoldi et al., 2007). We found 
there are 9 distances $< 60''$ between the 71 sources of these three fields, and so  $f=9/(71-3)=0.13$.
The factor (71-3) is the number of nearest neighbour distances for 71 sources distributed in the 3 separate fields.  
For the five sources around GJ526 in Fig~1, all 5-1  nearest neighbour distances are $< 60''$ as already mentioned,  so $k=N=4$ and  the probability $B_k$ 
that background sources can produce such a number of small distances is as low as 0.1\%.



\begin{table*}
\caption{New MAMBO-2 observations at 1.2~mm for nearby M-dwarfs. The table includes mean brightness determined
 by averaging map intensities over an effective area to  search for faint debris disks that are not readily apparent.
Not shown in this table are the  five emission clumps symmetricly located around GJ526 that  might be an inclined  debris disk 
(see subsection 4.1.3)}
\begin{tabular}{l|lccccccc} \hline\hline\\\
       Name   & Spectral &   D     & Gal. l.      & Integration &  Map rms           &   Mean  brightness         & $\mu/\sigma_{\mu}$   &  $3\sigma$ flux density          \\
              & type     &  (pc)   & ($^{\circ}$) & time     & (mJy/$11''$~beam)      &   $\mu \pm \sigma_{\mu}$   &        (2)           &  upper limit (1.2~mm)            \\
              &          &         &              & (hours)  &                       &    (mJy/$11''$~beam)        &                      &   (mJy)                          \\
              &          &         &              &          &     (1)               &        (2)                 &                      &        (3)                       \\\\
\hline\\\
  GJ699       &  M4.0  &   1.82  &   14  &   1.5  &   1.22  &  -0.03 $\pm$ 0.08  &  -0.36  &  $<$ 7.9  \\  
  GJ406       &  M6.0  &   2.38  &   56  &   2.5  &   0.99  &  -0.18 $\pm$ 0.08  &  -2.29  &  $<$ 4.9   \\  
  GJ411       &  M2.0  &   2.54  &   65  &   6.5  &   0.67  &   0.06 $\pm$ 0.06  &   0.97  &  $<$ 3.1   \\  
  GJ905       &  M5.5  &   3.16  &  -17  &   0.5  &   2.30  &   0.09 $\pm$ 0.24  &   0.38  &  $<$ 8.6   \\  
  GJ447       &  M4.0  &   3.34  &   59  &   1.5  &   1.19  &  -0.09 $\pm$ 0.13  &  -0.69  &  $<$ 4.2   \\  
  GJ725(*)    &  M3+M3.5 & 3.57  &   24  &   1.0  &   1.44  &   0.29 $\pm$ 0.17  &   1.69  &  $<$ 4.7   \\  
  GJ1111      &  M6.5  &   3.62  &   32  &   1.0  &   1.75  &   0.09 $\pm$ 0.20  &   0.44  &  $<$ 5.7   \\
  GJ54.1      &  M4.5  &   3.72  &  -79  &   1.5  &   1.49  &  -0.33 $\pm$ 0.18  &  -1.86  &  $<$ 4.7   \\  
  GJ273       &  M3.5  &   3.79  &   10  &   0.5  &   2.13  &   0.17 $\pm$ 0.25  &   0.68  &  $<$ 6.6   \\
  GJ234       &  M4.5+M8 & 4.12  &   -6  &   2.5  &   1.03  &   0.30 $\pm$ 0.13  &   2.23  &  $<$ 2.9   \\
  GJ83.1      &  M4.5  &   4.44  &  -46  &   1.5  &   1.40  &  -0.37 $\pm$ 0.18  &  -2.11  &  $<$ 3.7   \\
  GJ687       &  M3.0  &   4.53  &   32  &   4.0  &   0.76  &   0.25 $\pm$ 0.11  &   2.18  &  $<$ 2.9  \\  
  LHS292      &  M6.5  &   4.54  &   41  &   0.5  &   2.29  &  -0.35 $\pm$ 1.52  &  -0.23  &  $<$ 5.9   \\  
  GJ1002      &  M5.5  &   4.69  &  -69  &   0.5  &   2.23  &   0.14 $\pm$ 0.33  &   0.43  &  $<$ 5.6   \\
  GJ412       &  M2+M6 &   4.83  &   63  &   0.5  &   2.26  &  -0.21 $\pm$ 0.36  &  -0.58  &  $<$ 5.4   \\  
  GJ388       &  M3.0  &   4.89  &   54  &   5.0  &   0.73  &   0.20 $\pm$ 0.06  &   3.25  &  $<$ 1.8  \\  
  GJ445       &  M3.5  &   5.38  &   38  &   1.0  &   1.55  &   0.05 $\pm$ 0.28  &   0.18  &  $<$ 3.4   \\
  LHS1723     &  M4.5  &   5.43  &  -27  &   0.5  &   2.23  &   0.49 $\pm$ 0.42  &   1.17  &  $<$ 4.8   \\
  GJ526(**)  &  M1.5  &   5.43  &   72  &  16.0  &   0.58  &  -0.02 $\pm$ 0.12  &  -0.16  &  $<$ 1.4   \\  
  GJ251       &  M3.0  &   5.57  &   15  &   1.0  &   1.54  &   0.14 $\pm$ 0.29  &   0.48  &  $<$ 1.7   \\  
  GJ205       &  M1.5  &   5.71  &  -19  &   1.0  &   1.86  &  -1.09 $\pm$ 0.37  &  -2.93  &  $<$ 2.0   \\  
  GJ213       &  M4.0  &   5.87  &   -9  &   0.5  &   2.19  &  -0.31 $\pm$ 0.44  &  -0.70  &  $<$ 2.4   \\  
  GJ908       &  M1.0  &   5.93  &  -57  &   0.5  &   2.01  &   0.24 $\pm$ 0.39  &   0.61  &  $<$ 2.2   \\
  GJ581       &  M3    &   6.27  &   40  &   0.5  &   1.96  &  -0.75 $\pm$ 0.43  &  -1.75  &  $<$ 2.1   \\  
  GJ625       &  M1.5  &   6.58  &   43  &   5.0  &   0.69  &   0.38 $\pm$ 0.15  &   2.53  &  $<$ 0.7   \\
  GJ408       &  M2.5  &   6.62  &   64  &   2.0  &   1.13  &   0.42 $\pm$ 0.14  &   3.05  &  $<$ 1.2   \\
  GJ109       &  M3.5  &   7.55  &  -31  &   3.0  &   0.98  &   0.32 $\pm$ 0.22  &   1.48  &  $<$ 1.1   \\  
  GJ102       &  M4    &   7.75  &  -32  &   0.5  &   2.20  &  -0.09 $\pm$ 0.60  &  -0.15  &  $<$ 2.4    \\  
  GJ4247      &  M4    &   8.96  &  -21  &   3.5  &   0.74  &   0.63 $\pm$ 0.25  &   2.56  &  $<$ 0.8   \\  
  GJ569 (*)   &  M2.5+M8.5 & 9.81  &  59  &  2.0  &   1.09  &  -0.46 $\pm$ 0.37  &  -1.25  &  $<$ 1.2   \\\\
\hline
\end{tabular}


\smallskip

\scriptsize 

{\bf (1)} rms estimated for $r < 60''$ in the map ; 

{\bf (2)} mean brightness and uncertainty of mean computed by averaging intensities over an effective disk of radius  60~AU ;

{\bf (3)}  the $3\sigma$ flux density upper limit is computed also over the same  effective area (radius=60AU).

{\bf (*)} : background sources in map (see Table 2)~;

{\bf (**)} : possibly a large debris disk (see Fig~1 and subsection 4.1.3). 

\end{table*}

\begin{table*}
\caption{M-dwarfs associated with moving groups observed  at 1.2~mm in survey~I with the MAMBO-2 facility and already presented in Lestrade et al. (2006) but reanalyzed  
here in a consistent fashion with the  new data of  Table 3.}
\begin{tabular}{l|lccccccc} \hline\\\\
       Name   & Spectral &   D     & Gal. l.      & Integration &  Map rms   &   Mean  brightness         & $\mu/\sigma_{\mu}$  &  $3\sigma$ flux density   \\
              & type     &  (pc)   & ($^{\circ}$) & time     & (mJy/$11''$~beam)    &   $\mu \pm \sigma_{\mu}$   &     (2)            &  upper limit (1.2~mm)  \\
              &          &         &              & (hours)  &               &    (mJy/$11''$~beam)        &                     &  (mJy)            \\
              &          &         &              &          &   (1)         &       (2)                  &                     &      (3)          \\\\
\hline\\\ 
  GJ65        &  M5.5  &   2.6  &  -76  &   1.0  &   2.30  &  -0.15 $\pm$ 0.17  &  -0.87  &  $<$ 10.4   \\  
  GJ628       &  M3.5  &   4.5  &   24  &  20.0  &   0.51  &  -0.06 $\pm$ 0.08  &  -0.74  &  $<$  1.3   \\  
  GJ876       &  M4    &   4.7  &  -59  &   0.5  &   2.80  &   0.17 $\pm$ 0.46  &   0.37  &  $<$  7.0   \\  
  GJ873       &  M3.5  &   5.0  &  -13  &   0.5  &   2.31  &   0.33 $\pm$ 0.38  &   0.86  &  $<$  5.4   \\  
  GJ3379      &  M4    &   5.4  &  -10  &   0.5  &   2.57  &   0.05 $\pm$ 0.56  &   0.09  &  $<$  5.6   \\  
  GJ285       &  M4.5  &   5.9  &   13  &   1.0  &   1.74  &  -0.88 $\pm$ 0.35  &  -2.52  &  $<$  1.9   \\  
  GJ402       &  M4    &   6.8  &   55  &   1.0  &   1.81  &  -0.86 $\pm$ 0.39  &  -2.18  &  $<$  2.0   \\  
  GJ393       &  M2    &   7.2  &   47  &   1.5  &   1.51  &  -0.50 $\pm$ 0.33  &  -1.50  &  $<$  1.6   \\  
  GJ849       &  M3.5  &   8.8  &  -45  &   0.5  &   2.71  &   0.66 $\pm$ 0.89  &   0.74  &  $<$  2.9   \\  
  GJ791.2     &  M4.5  &   8.9  &  -17  &   1.0  &   1.70  &  -0.87 $\pm$ 0.57  &  -1.53  &  $<$  1.8   \\
  GJ856A      &  M3    &  16.0  &  -20  &   1.1  &   1.50  &  -0.56 $\pm$ 1.50  &  -0.37  &  $<$  1.6   \\
  GJ9809      &  M0    &  24.0  &  03   &   0.5  &   2.20  &   0.36 $\pm$ 2.20  &  0.16   &  $<$  2.4  \\\\\hline
\end{tabular}

\smallskip

\scriptsize 

{\bf (1)} rms estimated for $r < 60''$ in the map ;
 
{\bf (2)} mean brightness and uncertainty of mean computed by averaging intensities over an effective disk of radius 60~AU ;

{\bf (3)}  the $3\sigma$ flux density upper limit is computed also over the same  effective area (radius=60AU).

\end{table*}

\begin{figure}[t]
\centering
\includegraphics[width=8.5cm, angle=-90]{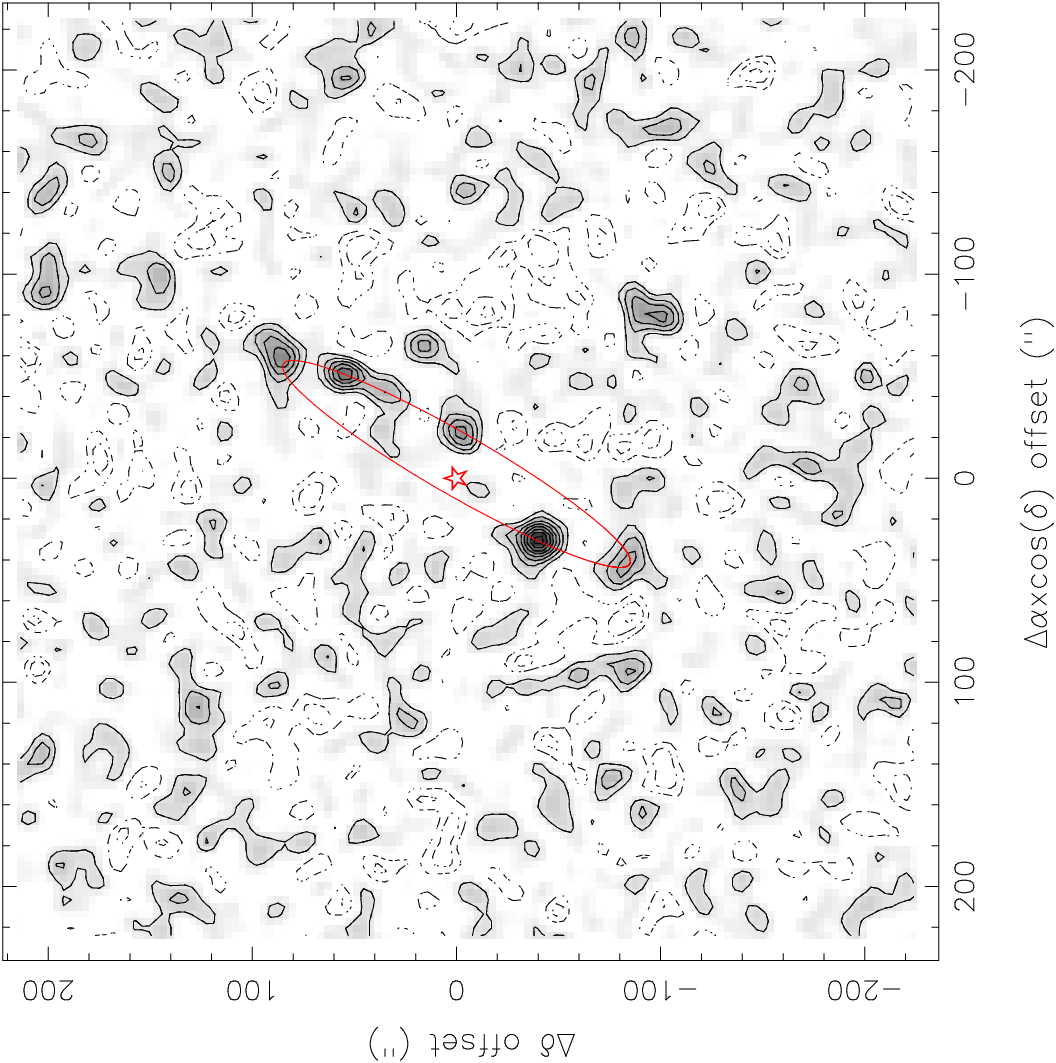}
\caption{MAMBO-2 Signal-to-Noise ratio map of   the field  around the  M1.5 dwarf GJ526 at $\lambda=1.2$~mm. 
The pixel size is $3.5'' \times 3.5''$. The noise  rms  is $ \rm \sigma = 0.58~mJy/11''$~beam 
in the central region ($r < 60''$) and increases towards the edges of the map, $\sim 1.3$~mJy/$11''$~beam at $r \sim 140''$.
The contours are $-4 \sigma,-3 \sigma, -2 \sigma, -1 \sigma$ (dotted lines), 
and $1 \sigma, 2 \sigma,   3 \sigma,  4 \sigma,  5 \sigma,  6 \sigma, 7  \sigma, 8 \sigma$.
The ellipse ($a=98''$, $b=15''$ and  PA=$-30^{\circ}$) is  almost centered  on the star position  and connects  five  sources
that might be  clumps of a debris disk  inclined to the plane of the sky by  $81^{\circ}$ (see subsection 4.1.3). 
The field is centered on the  position of GJ526 of early 2006 : $\alpha(J2000) =$ 13h 45m 44.52s  and $\delta(J2000)=$ $14^{\circ} 53' 20.6''$ 
(red star). See Table~2 for integrated flux densities and coordinates of  sources.}
\label{FigVibStab}
\end{figure}

\begin{figure}[t]
\centering
\includegraphics[width=7.0cm, angle=-90]{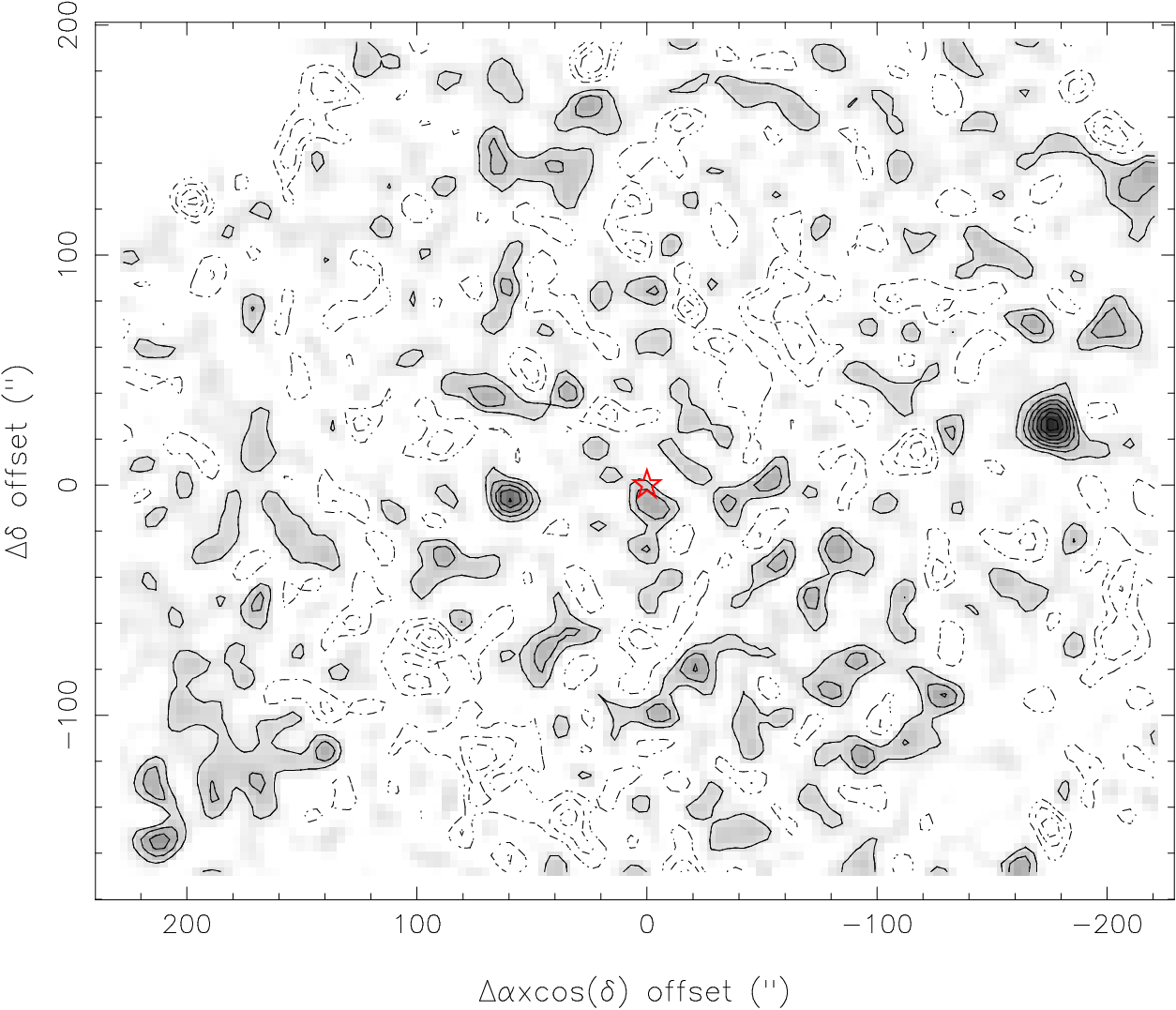}
\caption{MAMBO-2 Signal-to-Noise ratio map of the field around the binary  GJ725 (M3 + M3.5 dwarfs) at $\lambda=1.2$~mm. 
The pixel size is $3.5'' \times 3.5''$. The noise rms  is $ \rm \sigma = 1.4~mJy/11''$~beam 
in the central region ($r < 60''$) and increases towards the edges of the map,  $\sim 4.3 $~mJy/$11''$~beam at $r \sim 170''$.
The contours are $-4 \sigma, -3 \sigma, -2 \sigma, -1 \sigma$ (dotted lines), 
and $1 \sigma, 2 \sigma,   3 \sigma,  4 \sigma, 5 \sigma,  6  \sigma, 7  \sigma$.  
The  source MM184222+593828  westward has an integrated flux density between 30 and 63 mJy ($S/N \sim 7$). 
The other source, MM184253+593756,  has an integrated  flux density of 9.7mJy and a $S/N$ of 5.1.
The field is centered on the  position of GJ725A of early 2006 : $\alpha(J2000) =$ 18h 42m 45.48s  and  $\delta(J2000)=$ $59^{\circ} 38'  01.7''$
(red star). See Table~2 for integrated flux densities and coordinates of the sources.}
\label{FigVibStab}
\end{figure}

\begin{figure}[t]
\centering
\includegraphics[width=7.0cm, angle=-90]{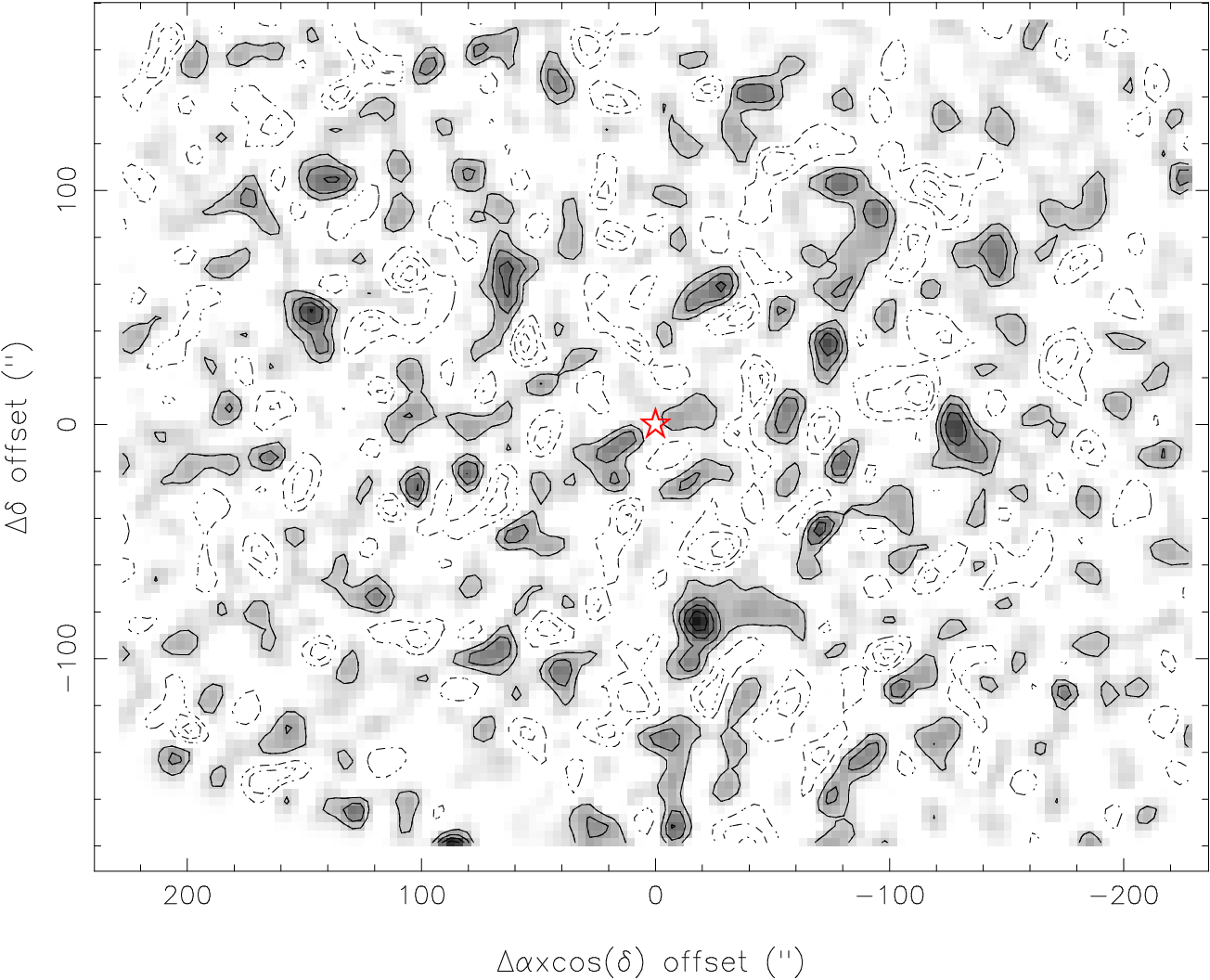}
\caption{MAMBO-2 Signal-to-Noise ratio map of the field around the binary GJ569 (M2.5 + M8.5 dwarfs) at $\lambda=1.2$~mm. 
The pixel size is $3.5'' \times 3.5''$. The noise  rms  is $ \rm \sigma = 1.1~mJy/11''$~beam 
in the central region ($r < 60''$) and increases  towards the edges of the map, $\sim 2.8$~mJy/$11''$~beam at $r \sim 140''$.
The contours are $-4 \sigma, -3 \sigma, -2 \sigma, -1 \sigma$ (dotted lines), 
and $1 \sigma, 2 \sigma,   3 \sigma,  4 \sigma$.
The  source MM145428+160439  has an integrated flux density of 5.3~mJy and a $S/N$ of $4.9$.  
The field is centered on the position of GJ569A of early 2006 : $\alpha(J2000) =$ 14h 54m 29.35s and $\delta(J2000)=$ 
$16^{\circ} 06' 03.2''$ (red star).
See Table~2 for integrated flux density and coordinates of the source.}
\label{FigVibStab}
\end{figure}

As a second test,  we use the elliptical pattern connecting  6 submm sources recognisable in  the North-West part of the COSMOS
field  mapped by MAMBO-2 and displayed  in Fig~2 of Bertoldi et al. (2007). This field is $20 x 20$~arcmin$^{2}$ in size and
the mean occurence of ellipse of major-axis $200''$, as in the field of GJ526, is $1/20 x 20$~arcmin$^{-2}$. 
We compute the Poisson probability to find a similar  ellipse  in a field that we take as small 
as $200 x 200$ arcsec$^{2}$  to account for the fact
that the ellipse around GJ526 is found  in an area restricted to the central part of  the map, {\it i.e.} centered
on the position of GJ525.  This probability is 2.7\%. This test is only indicative because the COSMOS field has been mapped
at the level of 1 mJy/beam rms while our map is twice as deep for the field around GJ526.  
 
In summary, the five  sources in Fig~1 that are symmetricly located around GJ526 are connected 
by an ellipse almost centered on the star. The two tests carried out above provide indications that this structure is statistically 
unconsistent with known spatial distributions of background submm galaxies.   
Hence,  at this stage, we cannot rule out  the hypothesis  that the sources are associated with the star.
In this case, the 5 connected sources could be indicative of  azimuthal structures in an inclined  debris disk around GJ526
whose extended emission is not seen because the map is not deep enough.   
Complementary observations at 850~$\mu$m, and at shorter wavelengths with Herschel 
should attempt to detect the extended emission of the disk. 
Eventually, astrometry should detect the same prope motion for the 5 connected sources as for the star GJ526 ($2.3''$/yr 
in the SE direction) if indeed they are part of a debris disk. The mid-epoch of our MAMBO-2 data is early 2007, so that the 5 sources 
should have moved in concert with the star by a full IRAM-30m beam by 2011, providing definitive proof of a disk.      

Such a debris disk would have a radius as large as $\sim 500$~AU at the distance of GJ526.
We examine whether or not this is conceivable. First, we note  that  exceptionally large debris disks,  
$520$~AU and 600~AU in extent, have recently been found  around the 184~Myr old A0-type dwarf $\gamma$~Oph at 70~$\mu$m (Su et al. 2008)
and around the main sequence F8V  $q^1~Eri$ at 870~$\mu$m (Liseau et al. 2008). Second, 
 protoplanetary disks where  planetesimals and planets form extend to almost 1000~AU 
 as observed for example around the young close binary  GG~Tau (Dutrey, Guilloteau \& Simon, 1994). 
In the model proposed by Kenyon and Bromley (2004a), icy-planets successively form in waves outward in the disk producing larger 
and larger dusty rings from collisional cascades. In their model, the planet formation timescale is  
$15-20 \times (\Sigma_0 / \Sigma_{MMSN})^{-1} (a/{\rm 30~AU})^3$~Myr in a quiet disk (their eq~(4)). 
Assuming for  $\Sigma_0$ as much as 15 times the  surface density of the minimum-mass  solar nebula  $\Sigma_{MMSN}$  as required for the formation of Jupiter  in the solar system (Lissauer 1987), 
the  timescale for planet formation to reach 500~AU in the GJ526 system is $\sim 5$~Gyr. 
The stellar diameter of GJ526 measured by the optical interferometer CHARA 
(Berger et al., 2006) is about $2\sigma$ larger than the ZAMS diameter predicted by Chabrier \& Baraffe (1997) and Siess et al. (2000). 
If such a  deviation of only $2\sigma$ is real, it is  an  indication of youth instead for GJ526, and this would pose a problem for 
the  Kenyon and Bromley model  applied  to  the hypothetical debris disk around GJ526. Note that, with the dust surface density profile  
$15 \Sigma_{MMSN} \times (r/r_0)^{-3/2}$, the mass of solids  in a ring at $r=500$~AU and $0.1 r$ in width is  $\rm 1~M_{\oplus}$ 
which is enough to form planetesimals.  Finally,   
the dust mass corresponding to the emission of the 5 sources is  between $\sim 6$ and  $\sim 10$ lunar masses 
as derived in the Appendix for dust grains  with a Dohnanyi size distribution 
and heated both by the stellar luminosity and the interstellar radiation field.
The dust mass is $\sim 22$ lunar masses  if  derived  conventionally 
for grey body dust with a  mass opacity  of $\rm 1.7 cm^2 g^{-1}$ at 850~$\mu$m and
$\propto \lambda^{-1}$ for  $\lambda > 210~\mu$m, and with the  single dust temperature 4.9K at large radius  from the star ($r=500$~AU) 
where the interstellar radiation field dominates the  heating process of grains (see Appendix).

\begin{table}

\caption{M-dwarfs observed at 850~$\mu$m with JCMT/SCUBA and already presented in  Lestrade et al. (2006)
but reanalyzed here in a consistent fashion with the new data of Tables 3 and 4.}

\begin{tabular}{l|lccrrrrcccc} 

\hline\hline\\

   Star    &   Sp.     &  Dist.   &  Integration &  Flux density  $^a$    & Size $^b$     \\
           &   type    &  (pc)    &  time        &  850~$\mu$m~~~~~~~      &  (AU)~~       \\
           &           &          & (hrs)        &   (mJy)~~~~~~~~        &               \\\\
\hline\\
 GJ82       &   M4      &   12.0    &   2          &    2.0  $\pm$ 1.4      &   84        \\
 GJ212      &   M0.5    &   12.5    &   2          &    1.3  $\pm$ 1.4      &   88        \\
 GJ285$^c$  &   M4.5    &    5.9    &   1          &    -0.7 $\pm$ 1.9      &   41        \\
 GJ393$^c$  &   M2      &    7.2    &   1          &    1.9  $\pm$ 1.9      &   50        \\
 GJ507.1    &   M1.5    &   17.4    &   1          &    -0.4 $\pm$ 2.0      &  121        \\
 GJ696      &   M0      &   21.9    &   2          &     0.8 $\pm$0.8       &  150        \\
 GJ9809$^c$ &   M0      &   24.9    &   1          &   -5.2  $\pm$ 2.3      &  174        \\
 GJ4247$^c$ &   M4      &    9.0    &   1          &    1.1  $\pm$ 2.1      &   63        \\
 GJ277B     &   M3.5    &   11.5    &   1          &   -2.1  $\pm$ 1.8      &   80        \\
 GJ842.2    &   M0.5    &   20.9    &   4          &   25.0 $\pm$ 4.6       &  300        \\
 GJ890      &   M2      &   21.9    &   2          &    -2.6 $\pm$ 1.6      &  153        \\
 GJ875.1    &   M3      &   14.2    &   1          &    0.52 $\pm$ 2.1      &   99        \\\\
\hline

\end{tabular}

\smallskip

\scriptsize
 
 $^a$  Flux density from  wide photometry.

 $^b$  Size is the radius of a debris disk as large as the telescope beam ($14''$).     

 $^c$  Stars also observed by MAMBO-2 in Tables 3 and 4.

\end{table}

\subsection{Deep search for debris disks in the MAMBO-2 maps}
  
We searched for faint debris disks  in each MAMBO-2 map by averaging intensities over a disk of increasing radius till $30''$ to see
whether or not  the mean brigthness  peaks at some angular radius $\theta$. The radius limit of $30''$ comes from the shift-and-add reduction method
and the wobbler throw used of $60''$ for the observations. The intention with this averaging was to find
a disk whose structure is not directly apparent in the map but whose  mean brightness is statistically significant. 
Naturally, no information on its structure can be recovered with this procedure. The optimum sensitivity of this method is
for face-on disks while  highly inclined disks might escape detections.   
This method is similar to the one used to determine extension limits of debris disks in 
the mid-IR surveys of Sun-like stars conducted by Smith, Wyatt \& Dent (2008), although,
we do not have the complication of having to accurately substract the photosphere at $\lambda=1.2$~mm.
Following this procedure, we plotted the mean  brightness as a function of 
angular radius $\theta$ for  each star of Tables~3 and 4.
All the curves were inspected and found to wander around zero mean with excursions $\le 3\sigma$ for $\theta$ comprised between $11''/2$ and $30''$, 
indicating that no new disk was found by this method. 

In Fig~4, we provide the distribution of the  {\sl mean brightness} $\mu$ / {\sl uncertainty 
of mean}  $\sigma_{\mu}$ listed in Tables~3 and 4 and computed 
for $\theta$ corresponding specificly to the adopted disk radius 60~AU at the distances of the stars.  
The corresponding Gaussian probability density function is also plotted in Fig.~4. 
Comparison between the two distributions indicates there are more 
high  positive and negative ratios $\mu / \sigma_{\mu}$ than expected although they stay  within $\pm 3\sigma$.
It means that there are still some systematic errors in the maps 
but at a low level, likely caused by remaining atmospheric fluctuations.

\subsection{Upper limits on fractional dust luminosities and on dust masses of the debris disks}

The fractional dust luminosity is the fraction 
of the stellar radiation  absorbed and reprocessed to the infrared and (sub)mm by the dust grains~; it is 
proportional to the fraction of the sky covered by dust as seen from the star (Dominik and  Decin 2003). 
We used the Stefan-Boltzmann  law $-$ black body emission $-$
to estimate  dust luminosity,  and modified it by emissivity $1/X_{\lambda}$, with   
$X_{\lambda} =1$ for $\lambda < 210~\mu$m and $X_{\lambda} = \lambda/210$ for $\lambda > 210~\mu$m (Wyatt 2008). 
In these conditions, fractional dust luminosity is~:

  $${L_{dust} \over L_*} = {{9 \times 10^{6} c^2} \over {2 \pi h \nu^3}}~(e^{h \nu/kT_d} - 1)~{{S_{\theta}~d^2~T_d^4} \over {R_*^2~T_*^4}} \times X_{\lambda}~~~~~~~~(1)$$ 
  
\noindent the normalization coefficient is such that the measured flux density  $S_{\theta}$ is in $Jy$, 
the star distance $d$  in $pc$, 
  the stellar radius $R_*$ in $m$, the dust and stellar effective temperatures $T_d$ and $T_*$  in $K$,
the Planck and Boltzmann constants $ h$ and $k$  in $J \times s$ and $J/K$, the speed of light $c$ in $m/s$, and the frequency
of observation $\nu$ in $Hz$. 

The standard argument used to fix dust temperature $T_d$ in mid-IR surveys of debris disks is that observations are most
sensitive to the dust emission  peak, and $T_d$
is derived from the Wien law. In the  (sub)mm range, this law  is
not appropriate for $T_d$ since we observe in the  Rayleigh-Jeans limit.   
To keep full generality, we avoid choosing a dust temperature  $T_d$ at some arbitrary radius  but   
plot in Figs~5 and 6  fractional dust luminosities and dust masses as functions of  disk radius $r$ 
comprised between 1 and 1000~AU following the approach by  Bryden et al. (2006) and Wyatt (2008).  
At disk radius $r$, we use the dust temperature :

  $$T_d=278 \times (L_*^{0.25}) \times (r^{-0.5})~~~~~~~~(2)$$   

\noindent from black body equilibrium where $L_*$ is the stellar luminosity
in $L_{\odot}$ and   $r$ is in AU (Backman \&  Paresce 1993). Note that by
combining eqs~(1) and (2), one gets the expression for  fractional dust luminosity given by
eq~(8) of Wyatt (2008) where the stellar luminosity $L_*$ cancels out. 

We adopt the $3 \sigma$  flux density limit for $S_{\theta}$ in eq~(1) by integrating brightness over a  face-on disk  
of radius $r$, or $\theta$ at the distance of the star. If $2 \theta > 11''$ ({\it i.e.}$>$~IRAM~30m~beam) : 
 
 $$S_{\theta} = 3 \times {rms} \times (2 \theta''/11'')  ~~~~~{\rm (mJy)} $$     

\noindent This formula  
takes into account that the mean brightness uncertainty $\sigma_{\mu}$ improves as
${rms}/\sqrt{number~of~beams}$  while the integrated flux density increases as 
$number~of~beams$ of the disk area.  Now if $2 \theta < 11''$~:

$$S_{\theta} = 3 \times {rms} ~~~~~ {\rm (mJy)} ~~~~~~~~~~~~~~~ $$
 
\noindent We have used the rms of Tables~3 and 4 that correspond to the central part of the maps, 
 thus underestimating slightly the upper limits computed.

In Fig.~5, we show the resulting fractional dust luminosity upper limits for $r$ comprised between 1 and 1000~AU. 
These functions first show a steep negative slope  as long as $2 \theta < 11''$ making  $S_{\theta}$ 
constant in eq~(1), then these functions level off when  $S_{\theta}$ linearly increases with $\theta$, finally they
increase when the dust temperature saturates at 4.9K because the  interstellar radiation field becomes dominant
over the stellar field (see Appendix). In this figure, we have added the upper limits of the fractional dust 
luminosities for the 41 M-dwarfs observed at 70~$\mu$m by Spitzer (Gautier et al. 2007),  
computed in a similar fashion from their $3\sigma$ flux densities. The figure shows that the two sets of data
are complementary, and lead to a uniform fractional dust luminosity  over a large extent  of disk radii for the
dozen of M-dwarfs  common to the two data sets.
In Fig.~6, we present the corresponding dust masses  as a function of $r$ for both samples computed with  the optically thin emission 
model ({\it e.g.} Zuckerman 2001) and  the mass opacity $\kappa_{\lambda} \propto \lambda^{-1}$ for  $\lambda > 210~\mu$m 
and   $\kappa_{850\mu m}  = \rm 1.7 cm^2 g^{-1}$.

\begin{figure}[t]
\centering
\includegraphics[width=6.5cm, angle=-90]{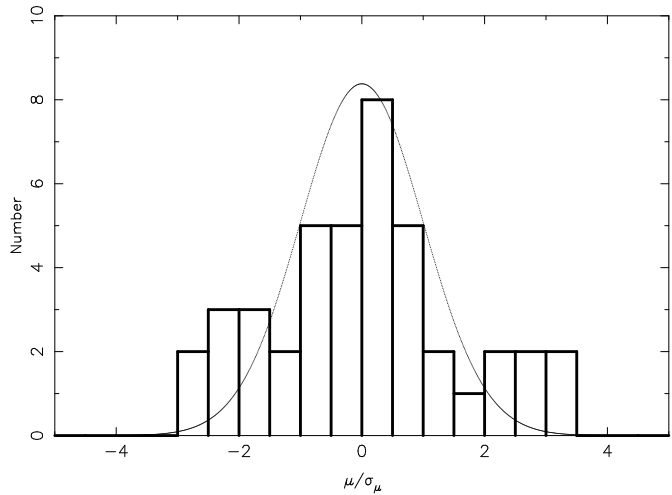}
\caption{Distribution of the ratios {\sl mean brightness}~$\mu$~/~{\sl uncertainty of mean}~$\sigma_{\mu}$ for the 42 M-dwarfs observed 
with MAMBO-2,  {\it i.e.}   $\mu / \sigma_{\mu}$ listed in Tables~3 and 4. Here, the  mean brightness  $\mu$  is computed 
over a disk of  radius 60~AU adopted for each star. The Gaussian probability
density function superimposed is scaled so that its integral is 42. Comparison between the two distributions indicates there are more 
high  positive and negative ratios $\mu / \sigma_{\mu}$ than expected although they stay  within $\pm 3\sigma$.
It means that there are still some systematic errors in the maps but at a low level, likely caused by remaining atmospheric fluctuations. 
No  mean brightness is retained as statistically significant in the sample.}
\label{FigVibStab}
\end{figure}

\begin{figure}[t]
\centering
\includegraphics[width=6.3cm, angle=-90]{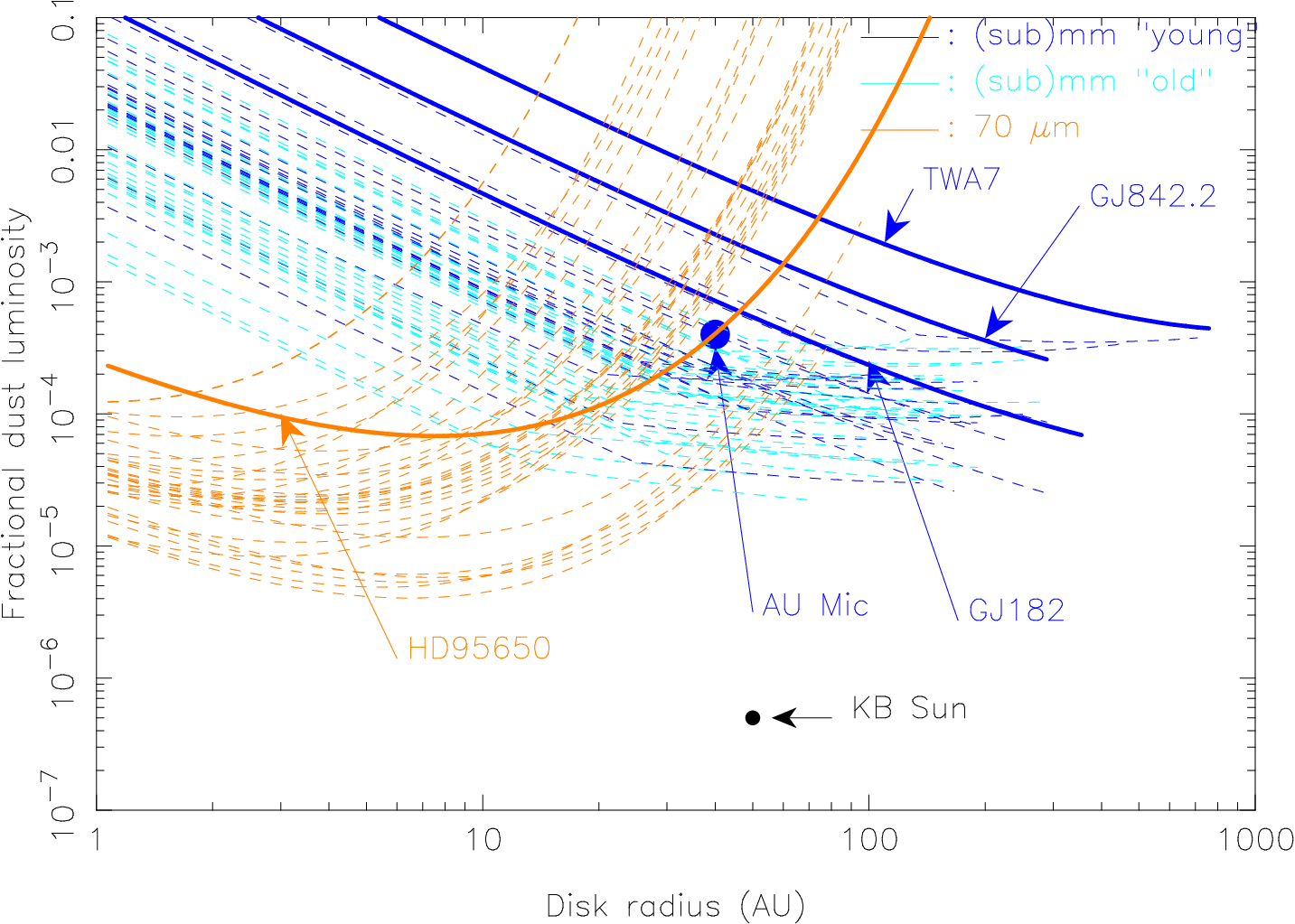}
\caption{
Constraints for the  dust luminosity fraction versus disk radius.
Note that the x-axis shows the single radius corresponding
to the single temperature of a disk (ring) assumed infinitely narrow in our model. We consider the wide range of  radii from
1 to 1000 AU as plausible for rings of debris. 
Upper limits are shown as dashed lines, detections as solid lines. 
\underline {Blue lines} correspond to the (sub)millimeter observations~; {\it i.e.} our sample of 50 M-dwarfs for which  dark blue is 
used for ''young'' M-dwarfs (ages $<$ 200 Myr) 
and  light blue  for ''old'' M-dwarfs (likely a few Gyr)~;  the two submm disk detected around AU~Mic 
and resolved (Liu et al., 2004, Liu 2004)  marked by a single blue dot~; the submm 
disks detected but not clearly resolved around GJ182 (Liu et al. 2004) and around GJ842.2 
(Lestrade et al. 2006)~; the submm transition disk 
detected around  the pre-main sequence M1 dwarf TWA7 (Matthews, Kalas, \& Wyatt, 2007) was also included.  
\underline {Orange lines} correspond to  Spitzer 70~$\mu$m observations~;  {\it i.e.} the  sample of 41 M-dwarfs  
of Gautier et al. (2007) and  the detection of the M0 dwarf HD95650 (Smith et al. 2006).
There are  16 M-dwarfs in common between the two samples. These two sets of data are complementary to
constrain  the existence of warm dust around M-dwarfs at moderate radii ($<$ 20 AU) and cold dust at large radii  ($>$ 20 AU).  
Disk diameters probed by the observations are limited by the angular size  of $60''$ for the MAMBO-2 maps, 
of   $28''$ for SCUBA wide photometry and of $38.4''$  for Spitzer aperture photometry (Gautier et al. 2007). 
Some curves are terminated at less than 1000 AU  because of these angular limits. 
The calculation of  the fractional dust luminosity is described in the text. (This figure is available in color 
in electronic form). }
\label{FigVibStab}
\end{figure}

\begin{figure}[t]
\centering
\includegraphics[width=6.3cm, angle=-90]{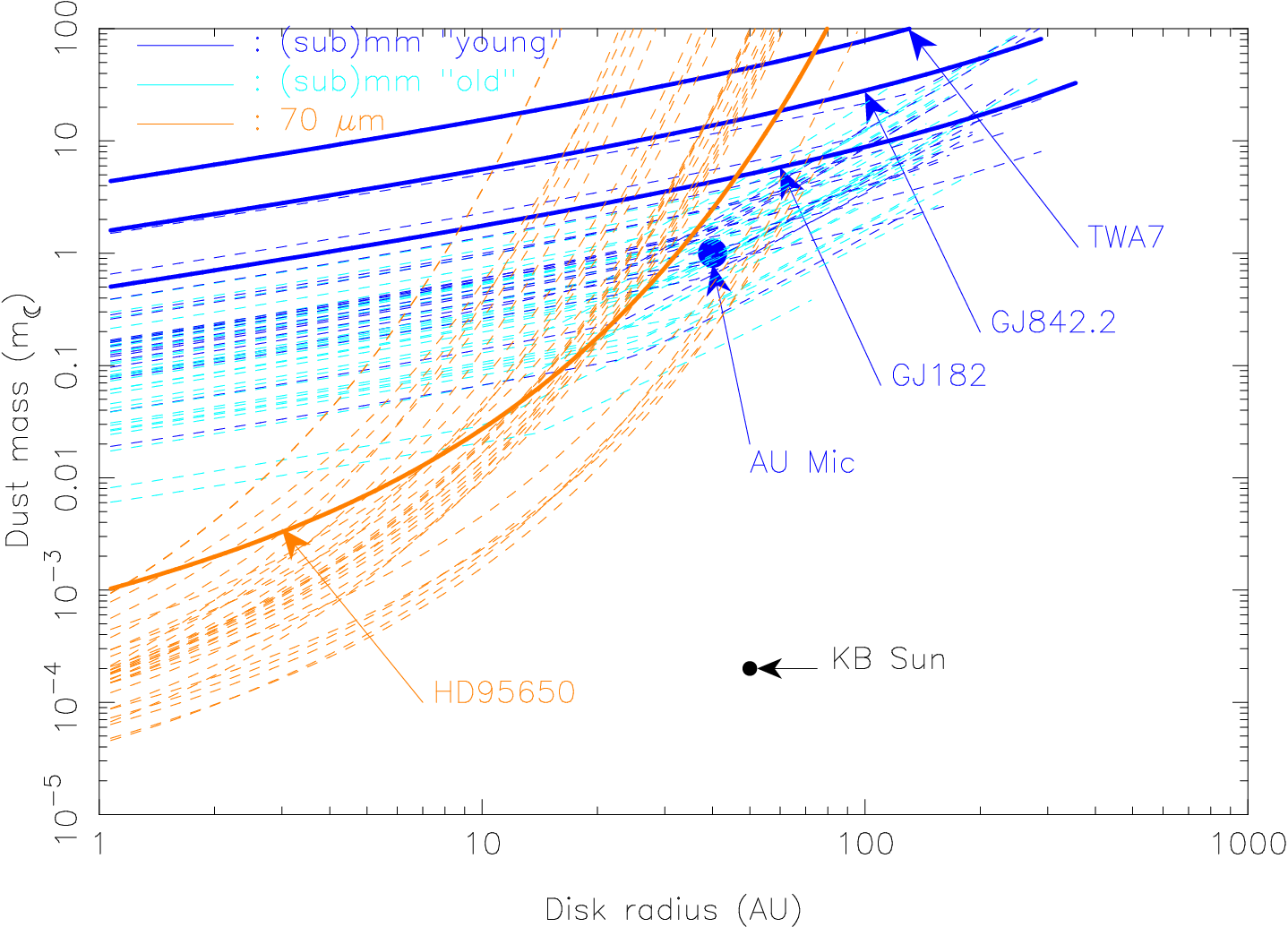}
\caption{Constraints for the dust mass versus disk radius. 
Note that the x-axis shows the single radius corresponding
to the single temperature of a disk (ring) assumed infinitely narrow in our model. We consider the wide range of  radii from
1 to 1000 AU as plausible for rings of debris. 
Upper limits are shown as dashed lines, detections as solid lines.
Blue lines correspond to the (sub)millimeter observations and
orange lines to  the Spitzer 70~$\mu$m observations  as detailed in the legend of Fig.5.
The mass opacity used to convert flux density to dust  mass with the standard optically thin emission model is
$\rm \kappa =  1.7 cm^2 g^{-1}$ at 850~$\mu$m and $\propto \lambda^{-1}$ for  $\lambda > 210~\mu$m. 
Additional information are  in the text. (This figure is available in color 
in electronic form).}
\label{FigVibStab}
\end{figure}

\begin{figure}[t]
\centering
\includegraphics[width=6.3cm, angle=-90]{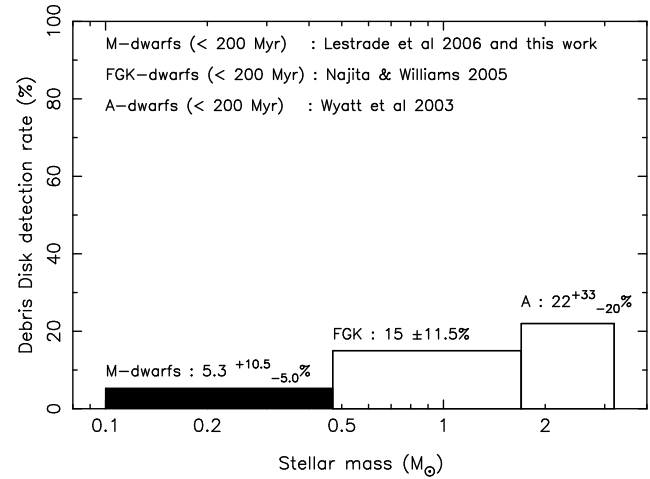}
\caption{
Detection rates of cold debris disks versus stellar masses (stellar types) for stars younger than 200~Myr.
These rates and uncertainties are for disks having dust fractional luminosities larger than the limits  
shown in Fig. 8 of the (sub)mm surveys used.}
\label{FigVibStab}
\end{figure}

\begin{figure}[t]
\centering
\includegraphics[width=6.3cm, angle=-90]{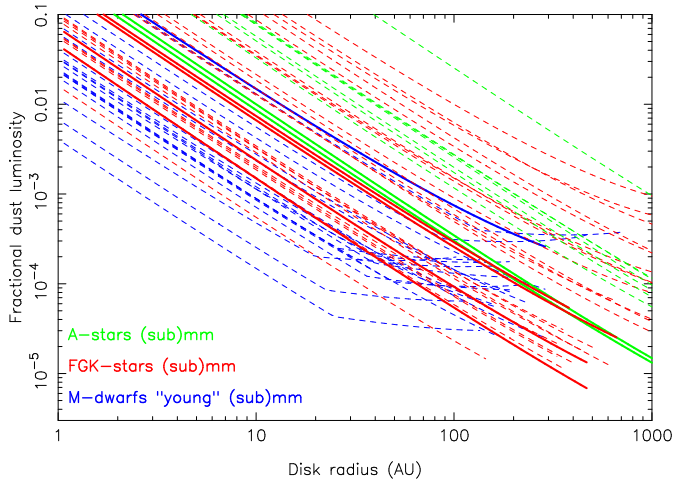}
\caption{
Dust luminosity versus disk radius limits of the three (sub)mm surveys used to determine  
the cold disk fractions of ''young'' M-, FGK- and A-type stars.
Shown are  the  dust luminosity fractions of the 19 youngest M-dwarfs of our sample (blue), 
of the 9 young A-stars (green) and of the 26 young FGK-stars (red) in the  two 850$\mu$m  surveys of Wyatt, Dent \& Greaves (2003)  
and of Najita \& Williams (2005).  All these stars are less than 200~Myr old. 
Most of the curves are upper limits (dashed lines).
The 7 detections are marked by full lines. This plot emphasizes  that for stars of lower masses (later stellar  types), 
even though the surveys are more sensitive, fewer disks are detected. 
Note that the x-axis shows the single radius corresponding
to the single temperature of a disk (ring) assumed infinitely narrow in our model. We consider the wide range of  radii from
1 to 1000 AU as plausible for rings of debris. 
(This figure is available in color 
in electronic form).}
\label{FigVibStab}
\end{figure}

\section{Discussion}

The single cold debris disk found in our surveys (GJ842.2, see details in Lestrade et al. (2006), and
excluding GJ526 at this stage)  makes the detection fraction to amount to  $2^{+4.5}_{-1.5}~\%$ 
in our sample of 50 M-dwarfs.   The limits  are  based on the Binomial distribution
for a small number sample and are such that  68\% of the probability is between the lower and upper uncertainties
and the peak probablity is  the observed fraction 1/50, following Burgasser et al. (2003).  
We recall that Gautier et al. (2007) had no detection in a sample of  62  nearby M-dwarfs  at 24$\mu$m  and
no  detection in a subsample of 41 of them at 70~$\mu$m, {\it i.e.} rate $<$ 7\%. 
Differently,  Forbrich et al. (2008) have detected photometric excesses at 24$\mu$m 
tracing warm dust around 11 M-dwarfs in the young open cluster NGC2457 (20-40~Myr) that represent $4.9 ^{+1.8}_{-1.8}~$\% 
of the 225 highly probable member M-stars identified in it. 

To discuss how  detectability of cold debris disks depends  
on the mass of the central star, we have compared our result with the observed fractions
 of cold debris disks around stars more massive than M-dwarfs. In the literature,
we find $\sim$ 30  cold  debris disks around  A-to-K type stars detected by submm observations
 (see Fig~3 and caption of Wyatt 2008). But most of these detections come from JCMT/SCUBA surveys
of IRAS biased samples, {\it i.e.} targets with prior IRAS excess detections, unlike our M-dwarfs, 
for which no such prior knowledge was used in  the selection of the sample.   
There are only two submm surveys of  A-to-K type stars that are unbiased in this respect  and that have
 depths of $\sim$1-3 mJy at 850~$\mu$m, comparable to our sensitivity at 1.2~mm~: 
the JCMT survey  by Najita \& Williams (2005) of thirteen   F5-to-K3 stars (10~Myr $<$ ages $<180$~Myr, $10 <$ d $< 78$~pc, 3 detections) and
the JCMT  survey by Wyatt, Dent \& Greaves (2003)  of nine  B7-to-A0 stars ($86<$ d $< 938$~pc, 2 detections) and of thirteen F3-to-K5 stars 
($27<$ d $< 250$~pc, 1 detection) that are all part of Lindroos binaries (14~Myr $<$ ages $<170$~Myr).
Combining these two submm surveys, we determine the cold disk fractions  of $22^{+33}_{-20}~\%$ for young A stars (2 detections/9) 
and of $15 ^{+11.5}_{-11.5}~\%$ for young FGK stars (4/26). For a meaningful comparison,  instructed by 
Spitzer surveys where the age factor is crucial for detectability of warm debris disks, we selected
the 19 youngest M dwarfs which have ages  between 20 and 200~Myr in our sample. The disk fraction  for these 
''young'' M-dwarfs is $5.3^{+10.5}_{-5.0}~\%$. Therefore, for this age range,  
there is an apparent trend in these three fractions, indicative of fewer cold disks 
detected for later stellar types $-$ lower star masses  $-$ although 
at a low statistical significance (Fig~7). Nonetheless this trend is notable because the surveys are deeper for later stellar 
types as shown in Fig~8, for  disk radii $<~100$~AU. Interestingly, this trend has recently been found also at $70\mu$m in a sample
of A to M stars with ages between 8~Myr and 1~Gyr by Plavchan et al (2009). 

We also determine the cold disk fraction of $< 10~\%$ for the ''old'' M-dwarfs of our sample having undetermined ages and likely being
as old as the average Galactic disk stars ($8.8 \pm 1.7$~Gyr, del Peloso et al. 2005). This may   
indicate that the cold disk fraction may decrease with stellar age, similarly to warm disks (Rieke et al. 2005).
We note that the ''old'' M-dwarfs have been observed with higher sensitivity as seen in Figs~5 and 6
where ''young'' and ``old'' M-dwarfs have been color-coded in  dark and light blue.  
The higher sensitivity for the ''old'' M-dwarfs is because they are nearer than the ''young'' M-dwarfs.

We now examine possible  reasons  why M-dwarfs have less detectable disks.
 
Theoretical models of star formation predict that  {\sl protoplanetary disks} are  less massive around M-dwarfs 
than around higher mass stars. Vorobyov \&  Basu (2008) predict  $<M_d>~\propto~<M_*>^{1.3\pm0.1}$, where disk mass $<M_d>$ and star
mass $<M_*>$ are time-averaged over the star accretion period (0.5 to 2.5 Myr). 
Consequently,  less primordial materials could  limit planet formation around M-dwarfs.  
Collecting  masses of protoplanetary disks determined by (sub)mm observations in the nearest star forming regions, 
Natta, Grinin, \& Mannings, (2000) found    $M_d~\propto~M_*^{0.64\pm0.14}$,
but Andrews \& Williams (2005, 2007) show that  $M_d$  versus $M_*$  in Taurus-Auriga is so widely scattered
between 0.001 and 0.2$M_{\sun}$ that it   
precludes any meaningful correlation fit. Also,  comparison of disk masses of 6 members of the
nearby young TW Hydrae Association (TWA)  suggests no correlation between disk masses and stellar
types for these reasonably coeval disks (Matthews, Kalas \& Wyatt 2007). 

{\sl Removal of  circumstellar dust} by the Poynting-Robertson effect and radiation pressure processes 
are diminished around  M-dwarfs because they are less luminous than solar-type stars, 
and so dust generated by collisions in any remnant planetesimal belt 
should remain there longer, giving rise to detectable emission.
However, the opposite conclusion has been reached by  Plavchan, Jura \& Lipscy, (2005) that highlight the
fact that  dust removal around M-dwarfs   could be dominated by the drag caused by strong   winds  associated with  
their  high coronal and chromospheric activities.

{\sl The formation of planetesimals and planets} depends on the 
time scales between the competing processes of coagulation and evaporation in the early period of accumulation. Theory predicts 
that  $t_{coag}$ increases and  t$_{evap}$  decreases with the central star mass and with the strength of the FUV and EUV radiation
field  (eqs~47 and 48 in Adams et al. 2004, respectively).  From their Fig~10, it can be seen  that
more than 10 Myr are needed to evaporate a protoplanetary  disk around a solar-mass star, whereas 
only a few Myr are  required to evaporate the same  disk around a low-mass M-dwarf 
in a stellar cluster with a  moderate UV flux of $\sim 3000G_0$. This might  quench  planet formation around  M-dwarfs.

 {\sl Early stripping of planetesimals} by passing stars  is likely since
most stars are born in clusters where stellar encounters as close as 160~AU are likely  in the first 100Myr 
(Kenyon  and Bromley 2004b). The disruption of  planetesimal disks by close stellar encounters
has been studied for the A6 star $\beta$~Pic by Larwood and Kalas (2001).
They found that, depending on the passing star's trajectory and on the relative star masses,   
1\% to 48\% of the planetesimals are lost after  encounters. Related issues for planetary systems were also discussed 
by Malmberg et al. (2007).

{\sl A Lack of gaseous giant planets} around M-dwarfs is  predicted by Laughlin, Bodenheimer \& Adams (2004),
caused by the longer duration required to build a core of 15$M_{\oplus}$. 
This lack of  giant planets might reduce the production of second generation dust in M-dwarf debris disks 
because of weaker gravitationnal stirring (Wilner et al. 2002),  diminishing their detectability. 

This series of arguments leads to the expectation that  debris disks around  M-dwarfs
might be intrinsically less dusty and therefore more difficult to detect than those around more massive stars of the same age.
However, most of these arguments   assume
that the initial protoplanetary  disk mass scales with the central star mass,
which may be plausible but is observationally not well established.

\section{Conclusion } 

To search for emission from cold debris disks, we have used the MAMBO-2 bolometer camera 
at the IRAM 30m telescope to map 42 nearby M-dwarfs at 1.2 mm wavelength to a noise level 
of 0.6 to 2.8 mJy per 11" beam. We also reanalyzed our earlier MAMBO-2 and SCUBA data 
to form a coherent sample of 50 M-dwarfs.
Only one cold debris disk was detected, surrounding the M0.5 dwarf GJ842.2. 
In an attempt to discuss how  detectability of cold debris disks depends  
on the mass of the central star, we have compared this result to the observed fractions of cold disks
for more massive stars in the two submm surveys  of Wyatt et al. (2003) and
Najita \& Williams (2005), who report detection rates of $22^{+33}_{-20}~\%$ for A-stars
and $15 ^{+11.5}_{-11.5}~\%$ for FGK-stars with stellar ages between 10 and 180~Myr. For
the 19 youngest M-dwarfs ($\le$~200~Myr) of our sample, we  found  a detection rate of $5.3^{+10.5}_{-5.0}~\%$. 
Hence, for this age range, there is a mild trend in these three detections rates, 
indicative of fewer cold debris disks detected for later stellar types $-$ 
lower star masses $-$ although at a low statistical significance. Nonethless this trend 
is notable because  the  sensitivities of these surveys are deeper for later stellar types.  
We  also determine the cold disk fraction of $< 10~\%$ for the ''old'' M-dwarfs (likely a few Gyr) of our sample,  
indicative that the cold disk fraction may decrease with stellar age, as is also  seen for warm disks.
Future observations of a larger and better controlled 
sample of stars of all stellar types  with Herschel in the far-IR and deeper observations 
in the (sub)mm will be able to better clarify these issues.
 
\section{Appendix : computation of the dust temperature and mass for the possible large debris disk around GJ526}

The source that heats  dust in debris disks is usually the stellar radiation field, but in a large debris disk, 
as possibly found around GJ526, the interstellar radiation field 
dominates at some radius  from the star. 
We compute  this transition radius  for several stellar spectral types  by  solving numerically the  integral equation for a grain at thermal equilibrium 
absorbing both the stellar and interstellar incident fields :

$$\int_0^{+\infty} Q_{abs}(\lambda,a)~.~\big( \pi a^2 J_*(\lambda,r) + 4 \pi a^2 J_{ISRF}(\lambda)  \big)~d\lambda = ~~~~~~~~~~~~~~~~~~~~~~~~~~~$$
$$\int_0^{+\infty}  4 \pi a^2  Q_{abs}(\lambda,a) \pi B(\lambda,T_g) d\lambda ~~~~~~~~~~~~~~~~~~~~~~~~~~~~~~~~~~~~~~~~~(A1)~~~~~~~~~~~~~~~~~~~~$$  

\begin{figure}[t]
\includegraphics[width=6.cm, angle=-90]{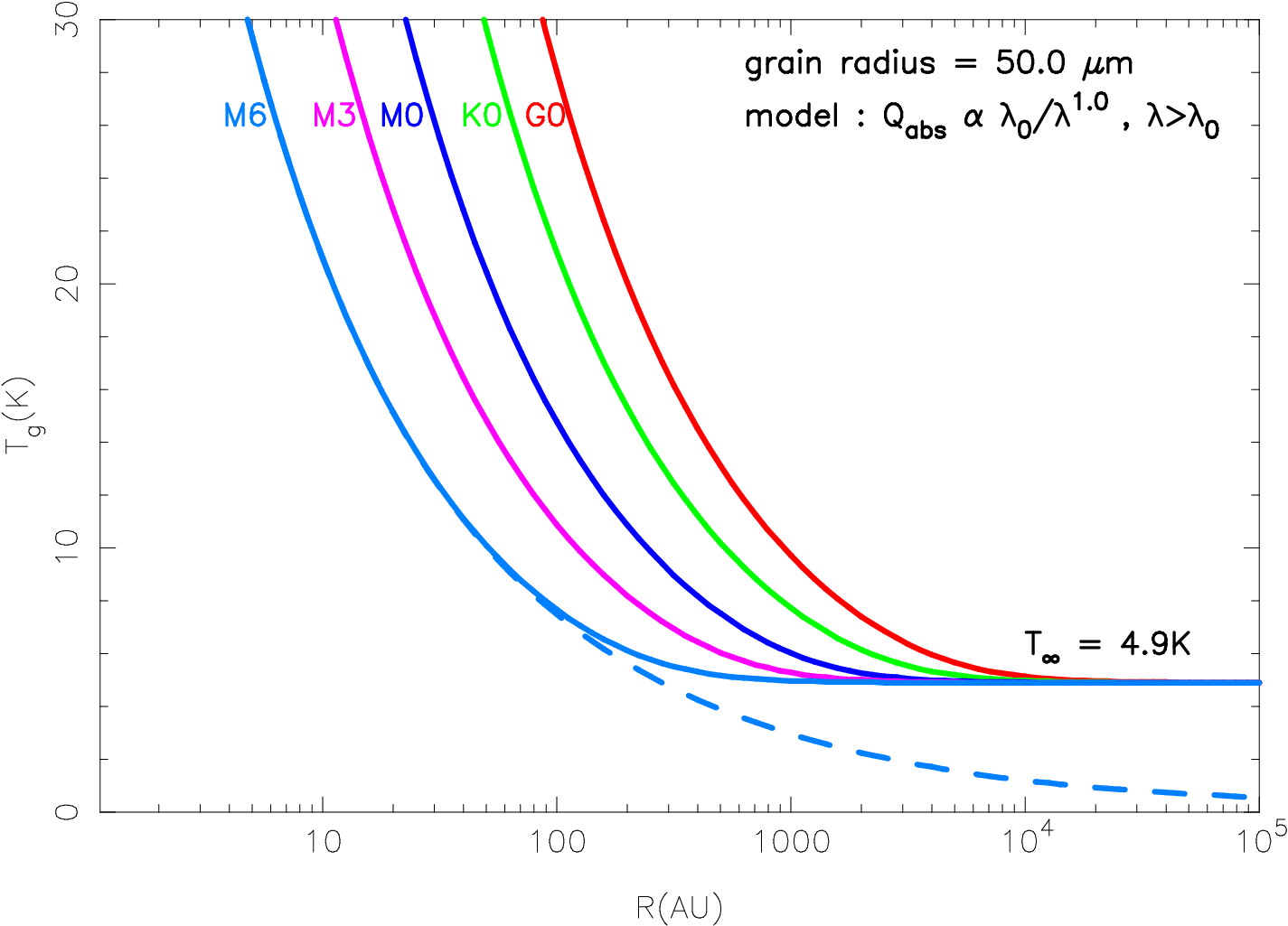}

{\bf Fig~A1} : Variation of the temperature of a grain 100$\mu$m in diameter exposed to the
radiation of the stellar field and of the isotropic interstellar radiation field. This latter field dominates 
at some disk radius  $R$ that depends on the stellar luminosity. The saturation temperation of 4.9K due to the   
interstellar field is consistent with the computation 
of Kr\"ugel (2003, p. 249). For M6 only, the dashed line indicates the grain temperature when the interstellar radiation field is not included.
(This figure is available in color in electronic form).
\end{figure}

\noindent where  $Q_{abs}(\lambda,a)= 1$ for  $\lambda < \lambda_0$ and   $Q_{abs}(\lambda,a)=  \lambda_0/\lambda$   
with $\lambda_0 = 2 \pi a$ for  $\lambda > \lambda_0$,  which approximate the absorption efficiency computed 
for carbon  by Laor \& Draine (1993). The parameter $a$ is the radius of a spherical grain.
The expression $B(\lambda,T_g)$ is the Planck function  at grain
temperature $T_g$ that depends on  $a$.  The intensity 
of  the stellar radiation field  $J_*(\lambda, r)$ is $ \pi B(T_*,\lambda) \times  {(4 \pi R_* /4 \pi r)}^2 $  at disk radius  $r$ 
with the star characterized  by its effective temperature T$_*$ and its radius R$_*$.  
Intensity of the Interstellar Radiation Field $J_{ISRF}(\lambda)$ in the solar neighbourhood is : 

$$J_{ISRF}(\lambda)= \sum_{i=0}^{i=6} C_i \pi B(\lambda, T_i) $$

\begin{figure}[t]
\includegraphics[width=6.cm, angle=-90]{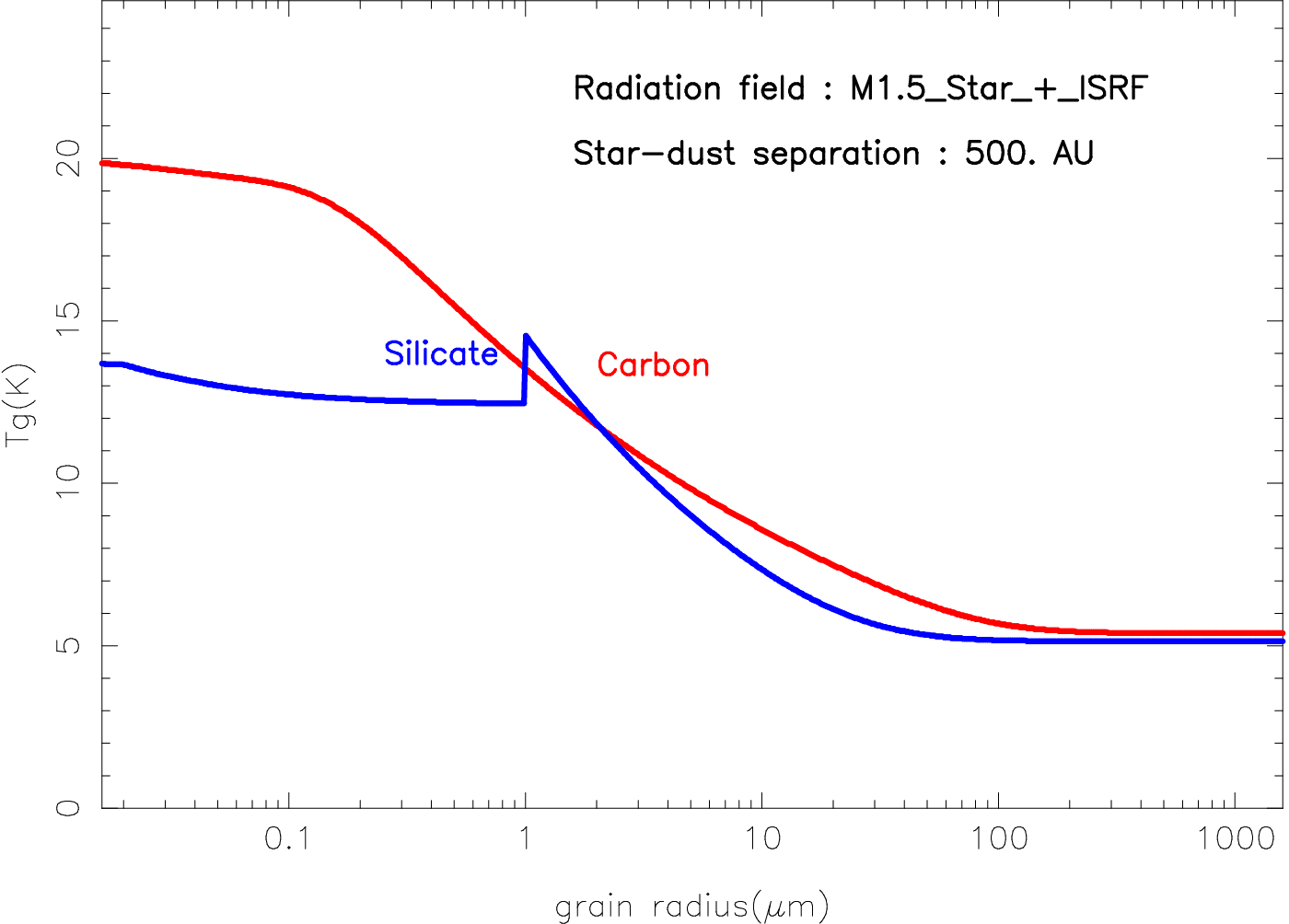}

{\bf Fig~A2} : Variation of temperature  as a function of grain radius at $500$~AU from the M1.5 dwarf GJ526.   
Similar results are reported by Kr\"ugel (2003, p. 249). The discontinuity in the silicate curve arises from the piecewise-defined funtion $Q_{abs}(\lambda,a)$ 
adopted from Laor \& Draine (1993). (This figure is available in color 
in electronic form).
\end{figure}

\noindent with components  $C_0=1, T_0=2.7K, C_1=4\times10^{-17}, T_1=27 000K,   C_2=10^{-14}, T_2=7500K,   C_3=10^{-13}, T_3=4000K,              
 C_4=4\times10^{-13}, T_4=3000K,  C_5= 1.5 \times 10^{-10}, T_5=400K, C_6= 0.5 \times 10^{-5}, T_6=46K$,
at galactocentric distance 10kpc (Mathis, Mezger \& Panagia 1983). 
In Fig~A1, we solved numerically eq~(A1) to determine dust temperature $T_g(r)$ as a function of disk radius   
in modelling grains with the  single size $2 a = 100 \mu$m, typical  
for submm observations,  and for 
 several stellar spectral types. For GJ526 (M1.5), this
figure shows that the interstellar radiation field becomes dominant at $r > 1000$~AU,
 but this transition radius is only 200~AU for spectral type M6. 

In Fig~A2, we use a more realistic model for the dust  in adopting the standard grain size
distribution $dN=N_0 a^{-3.5} da$ (Dohnanyi 1969) to compute  temperature  $T_g(a)$ as a function of grain
size  at $r=500$~AU with eq~(A1) but modified to include the grain size
distribution. Temperature increases significantly 
for $ a < 10~\mu$m which is relevant for M-dwarfs because the grain blow-out size is small. For GJ526 (0.031$L_{\odot}$), it is 
as small  as 0.04~$\mu$m for carboneous grains,  
ignoring the effect of stellar wind drag for this  star which  has a low coronal/chromospheric activity 
(Log$L_x$=26.87 [$10^{-7}W$], Schmitt \& Liefke, 2004).
Finally, we use this temperature function  $T_g(a)$  at  $r=500$~AU  to determine  the dust mass 
around GJ526 by matching  the total flux density
of the 5 clumps ($S_{\rm 1.2~mm}=  21.2 \pm 2$~mJy) to the flux density of our model. 
If $d$ is the distance to the star, the predicted flux density  is  :

$$ S_{\nu} = {N_0 \over 4 {\pi d^2}}~\int_{a_{min}}^{a_{max}} 4 \pi a^2~Q_{abs}(\nu,a)~.~\pi B(\nu,T_g(a))~.~a^{-3.5}~da ~~~~(A2)$$

\noindent the limit $a_{min}$ is set by the blow-out size ($a_{min}=0.04\mu$m for GJ526).
The limit $a_{max}$ is related to  the total dust mass  $M_d = {8 \over 3} \pi \rho N_0 \sqrt{a_{max}}$ for the size distribution above 
and  spherical grains of density $\rho$.   $M_d$   is the  dust  mass probed by the measured flux density at the observing $\lambda$.
Practically, the  grain size limit $a_{max}$ is  the value  that makes  convergent the computation of the power emitted by $M_d$  over 
a band $2b$ centered on the observed wavelength (1.2~mm). We adopted the convergence criterium  of 5\%  to match the relative accuracy 
of the measured flux density. In other words, $a_{max}$ is increased until  the integral below  converges to within 5\%  :

$$\int_{+1.2~mm - b}^{+1.2~mm + b} \int_{a_{min}}^{a_{max}} 4 \pi a^2~Q_{abs}(\lambda,a)~.~ \pi B(\lambda,T_g(a))~.~a^{-3.5}da~d\lambda ~~~~(A3)$$

\noindent This power would need to be computed more accurately by including the small contribution to the emission of 
larger pebbles only if the flux density were measured more accurately.        
We found that $a_{max}$ is  $\sim 24$mm and $\sim 36.4$mm for carboneous and silicate grains, respectively, 
and, independently of the choice  $b=5\%, 10\%, 20\%  \times 1.2$~mm. 
Similar calculation was made by Wyatt \& Dent (2002) for  the A3V dwarf Fomalhaut 
by integrating not only over grain size but also over a range of radius $r$ to account  for spatial distribution
of the dust in a ring. They found that 95\% of the flux density comes from grains and pebbles less than  100mm in radius. 
This is comparable to our determination of $a_{max}$ and the difference  must come from different function for $T_g(a)$ and  their   hypothesis of dust
spatially distributed. 
  
Figs~A3 and A4 show the SED of the dust around GJ526 based on our model for both carboneous and
silicate grains adjusted to the flux density measured at 1.2~mm. The
dust mass determined by our model  is between $\sim 6$ and   $\sim 10$ lunar masses. 
We have added in these figures, the SED  based on the standard grey-body model 
of  optically thin dust emission (Hildebrand 1983 and Zuckerman 2001) :

$$ S_{\nu}=~{M_d~\times~B(\nu,T_g)~\times~\kappa_{abs}}/{d^2} ~~~~(A4) $$

\noindent where  the   mass opacity
$\kappa_{abs}$ is conventionally taken to be  $\rm ~1.7~cm^2 g^{-1}$ at $850~\mu$m   and is scaled by  $\rm 210~\mu m / \lambda$
for  $\lambda > 210~\mu$m. This grey dust mass is about 2 to 4  times larger than the one determined by our model. This level of
accuracy  is typical  of dust mass estimates for debris disks, presently.  
     
The SED of our model is significantly more extended  to the Far-IR  and to the radio than  the grey-body model when both models match
the flux density at $\lambda=1.2$~mm. Noticeably, we  point out that this Far-IR extension is much less pronounced  
if our model is applied to an A-star.  
Future observations with Herschel will test these SED differences and thus will probe, in essence,  the collisional dust size
distribution adopted and the non-uniform grain temperature.

\begin{figure}[t]
\includegraphics[width=6.cm, angle=-90]{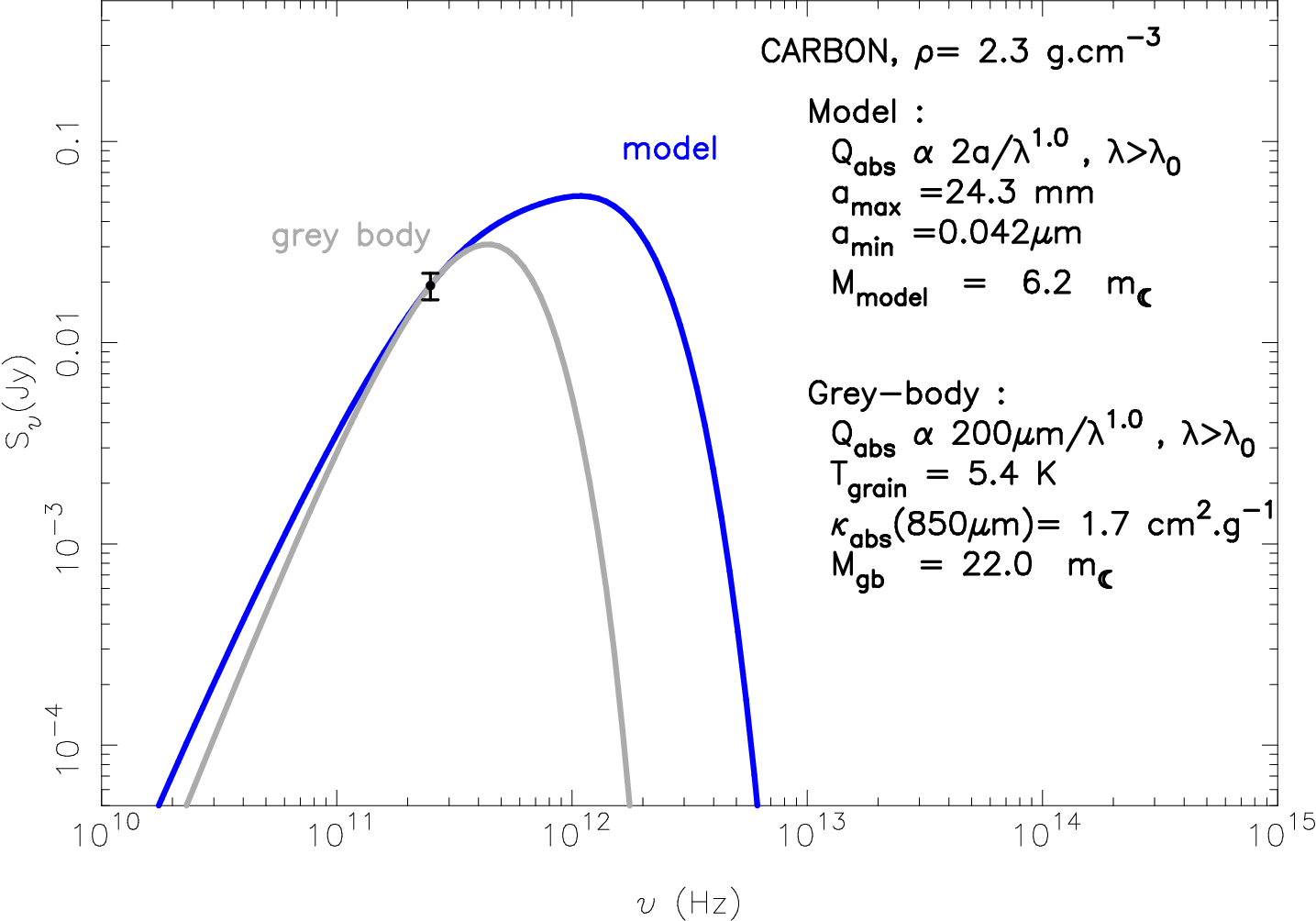}

{\bf Fig~A3} SED of  carboneous dust around GJ526 based on our model fitted to the 1.2~mm flux density.  
Our model includes the collisional dust size
distribution $dN=a^{-3.5}da$  and the resulting non-uniform  temperature for grains of various sizes computed in Fig~A2. 
Superimposed is the corresponding grey-body model computed with the single grain temperature 4.9 K and standard mass opacity 
$\rm 1.7~cm^2g^{-1}$ at $\lambda=850~\mu$m scaled with  the efficiency $\rm 210~\mu m / \lambda$ (see text).  
(This figure is available in color in electronic form).
\label{FigVibStab}
\end{figure}

\begin{figure}[t]
\includegraphics[width=6.cm, angle=-90]{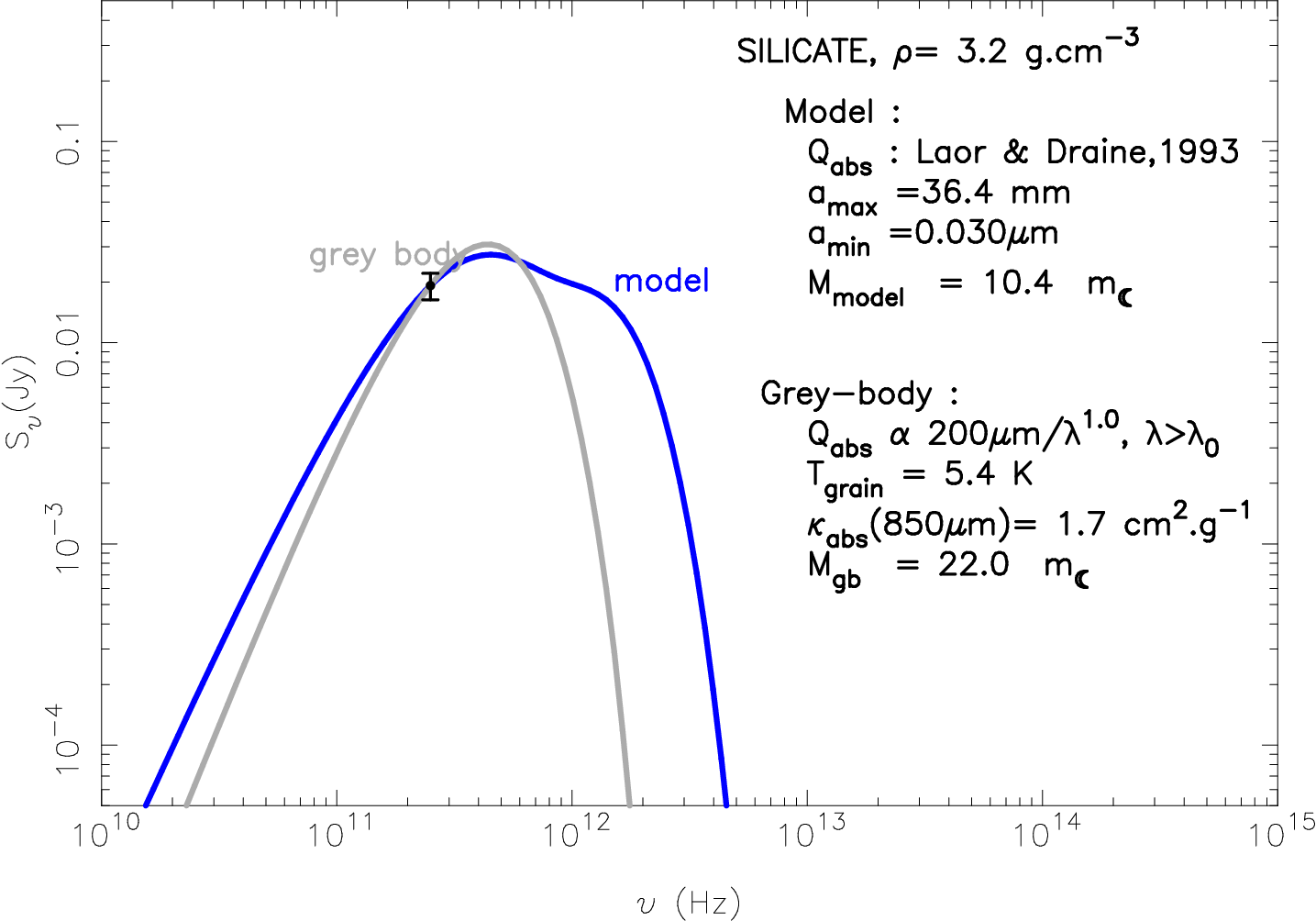}

{\bf Fig~A4}  SED of  silicate dust around GJ526 based on our model and fitted to the 1.2~mm flux density.  
Same comments  as in Fig~A3. (This figure is available in color 
in electronic form).
\label{FigVibStab}
\end{figure}

\begin{acknowledgements}

We are grateful to the staff of the IRAM 30-m telescope, especially Dr S. Leon,
for his dedication in managing the MAMBO-2 pool during the transistion period with the new contral system
and for his unfailing determination to optimize the science return of the telescope despite harsh conditions. 
We would like to thank an anonymous referee for his careful reading of the manuscript and constructive
remarks.
This research has made use of the SIMBAD database and  of the VizieR catalog access tool,
opera ted at the Centre de Donn\'ees Stellaires (CDS), Strasbourg,France.  
This publication makes use of data products from the Two Micron All Sky Survey, 
which is a joint project of the University of Massachusetts and
 the Infrared Processing and Analysis Center/California Institute of Technology, 
funded by the National Aeronautics and Space Administration and the National Science Foundation. 
This publication  has made use of data products from
the SDSS and SDSS-II funded by the Alfred P. Sloan Foundation, 
the Participating Institutions, the National Science Foundation, the U.S. Department of Energy, 
the National Aeronautics and Space Administration, the Japanese Monbukagakusho, the Max Planck Society, 
and the Higher Education Funding Council for England. 
This work is based in part on observations made with the Spitzer Space Telescope, 
which is operated by the Jet Propulsion Laboratory, California Institute of Technology under a contract with NASA.

\end{acknowledgements}

\null
\null

\noindent References 

\bibl
 Adams, F.C., Hollenbach, D., Laughlin, G., Gorti, U., 2004, ApJ, 611, 360.
\filbreak

\bibl
 Allard, F., Hauschildt, P. H., Alexander, D. R., Tamanai, A., \& Schweitzer, A.,  2001, ApJ, 556, 357
\filbreak

\bibl
 Andrews, S.M., \& Williams, J.P., 2005, ApJ, 631, 1134-1160
\filbreak

\bibl
 Andrews, S.M., \& Williams, J.P., 2007, ApJ, 671, 1800-1812
\filbreak

\bibl
Augereau, J.-C., \& Paploizou, J.C.B.,  2004, A \& A, 414, 1153
\filbreak

\bibl
Aumann, H. H., Beichman, C. A., Gillett, F. C., de Jong, T., Houck, J. R., Low, F. J., Neugebauer, G., Walker, R. G., Wesselius, P. R., 1984, ApJ, 278, 23
\filbreak

\bibl  
Backman D.E., Paresce, F., 1993, in Protostars and Planets III, ed. E.H. Levy and J.I. Lumine (Tucson : Univ. Press), 1253
\filbreak

\bibl 
Berger D.H.,  et al., 2006, ApJ, 644, 475 
\filbreak

\bibl 
Bertoldi, F.,  et al., 2007, ApJ Suppl., 172, 132 
\filbreak

\bibl 
Bryden, G. C., et al., 2006, ApJ, 636, 1098
\filbreak

\bibl 
Burgasser, A.J., et al., 2003, ApJ, 586,  512. 
\filbreak

\bibl 
Chabrier, G. \& Baraffe, I., 1997,  A\&A, 327, 1039
\filbreak

\bibl 
Condon, J.J., et al., 1998, AJ, 115, 1693
\filbreak

\bibl 
del Peloso, E.F., et al, 2005, A\&A, 440, 1153 
\filbreak

\bibl
Diggle, P.J., 2003, in {\it Statistical Analysis of Spatial Point Patterns}, Publisher : Hodder Arnold.
\filbreak

\bibl
 Dohnanyi, J.S., 1969, J. Geophys. Res., 74, 2531
\filbreak

\bibl
Dominik, C., Decin, G.,  2003, ApJ, 583, 626.
\filbreak

\bibl
Dutrey, A., Guilloteau, S., Simon, M., 1994, A\&A, 286, 149.
\filbreak

\bibl
Forbrich, J., Lada, C. J., Muench, A. A., Teixeira, P. S., 2008, ApJ, 687, 1107
\filbreak

\bibl
Gautier, T. N.,  Rieke, G. H., Stansberry, J., Bryden, G. C., Stapelfeldt, K. R., Werner, M. W., Beichman, C. A. et al., 2007, ApJ, 667, 527
\filbreak

\bibl
Gomes, R., Levison, H.F., Tsiganis, K., Morbidelli, A., 2005, Nature, 435, 466.
\filbreak

\bibl
Greaves, J. S., Holland, W. S., Wyatt, M. C., Dent, W. R. F., et al, 2005, ApJ, 619, 187
\filbreak

\bibl
Greve, T.R., et al, 2004, MNRAS, 354,779 
\filbreak

\bibl
Hahn, J.M., Malhotra, R., 1999, AJ, 117, 3041.
\filbreak

\bibl
Hildebrand, R.H., 1983, QJLR astr. Soc., 24, 267
\filbreak

\bibl
Holland, W.S., Greaves, J. S., Zuckerman, B., Webb, R. A., McCarthy, C.,  Coulson, I. M., Walther, D. M., Dent, W. R. F., Gear, Walter, K., Robson, I.,
1998, Nat., 392, 788-791
\filbreak

\bibl
Kalas, P., Graham, J.,  Clampin, M., 2005, Nature, 435, 1067 
\filbreak

\bibl
Kenyon, S.J., Bromley, B.C., 2004a, AJ, 127, 513-530
\filbreak

\bibl
Kenyon, S.J., Bromley, B.C., 2004b, Nature, 432, 598-602
\filbreak


\bibl
Kreysa, E., Gemünd, H. P., Gromke, J., et al. 1998, in Advanced Technology
MMW, Radio, and Terahertz Telescopes, ed. T. G. Phillips, SPIE, 3357, 319
\filbreak

\bibl 
Krist, J.E., et al., 2005,  AJ, 129, 1008
\filbreak

\bibl
Kr\"ugel, E., 2003,  in {\it The Physics of Interstellar Dust},  
Institute of Physics Publishing : Series Astronomy \& Astrophysics        
\filbreak

\bibl
Larwood, J.D., Kalas, P.G., 2001, MNRAS, 323, 402-416.
\filbreak

\bibl
Laughlin, G., Bodenheimer, P., Adams, F., 2004, ApJ, 612, L73-L76 
\filbreak

\bibl
Lestrade, J.-F., Wyatt, M. C., Bertoldi, F., Dent, W. R. F., Menten, K. M., 2006, A\&A, 460, 733 
\filbreak

\bibl
Liseau, R. et al., 2008, A\&A, 480, 47L
\filbreak

\bibl 
Lissauer, J.J., 1987, Icarus, 69, 249.
\filbreak

\bibl
Liu, M. C., Matthews, B.C.,  Williams, J.P., Kalas, P.G., 2004, ApJ, 608, 526
\filbreak

\bibl
Liu, M.C. 2004, Sci, 305, 1442
\filbreak

\bibl
 Laor, A. \& Draine, B.T., 1993, ApJ, 402, 441
\filbreak

\bibl
Malmberg, D. et al., 2007, MNRAS, 378, 1207
\filbreak

\bibl
 Mathis, J.S., Mezger, P.G. \& Panagia, N., 1983, A\&A, 128, 212
\filbreak

\bibl
Matthews, B.C., Kalas, P.G., Wyatt, M.C., 2007, ApJ, 663, 1103
\filbreak

\bibl
Monet, D.G., et al., 2003, AJ, 125, 984.
\filbreak

\bibl
Montes, D., et al., 2001, MNRAS, 328, 45
\filbreak

\bibl
Morbidelli, A., Levison, H.F., Tsiganis, K., Gomes, R.,  2005, Nature, 435, 462.
\filbreak

\bibl
Najita, J. and Williams, J.P., 2005, ApJ, 635, 625 
\filbreak

\bibl
Natta, A., Grinin, V., Mannings, V., 2000, Protostars and Planets IV (Book-Tucson : University of Arizona Press), p. 559
\filbreak

\bibl
 Plavchan, P., Jura, M., \& Lipscy, S.J., 2005, ApJ, 631, 1161
\filbreak

\bibl
Plavchan, P., Werner, M.W., Chen, C.H., Stapelfeldt, K.R., Su, K.Y.L., Stauffer, J.R., Song, I., 2009, astroph 0904.0819
\filbreak

\bibl
Reche, R.,  Beust, H., Augereau, J.-C., Absil, O., 2008, A\&A, 480, 551 
\filbreak

\bibl
Rieke, G. H., Su, K. Y. L.,  Stansberry, J. A., Trilling, D.,  Bryden, G.,  Muzerolle, J. et al, 2005, ApJ, 620, 1010
\filbreak

\bibl
Schmitt J.H.M.M., Liefke C., 2004, A\&A, 417, 651
\filbreak   

\bibl
Siess L., Dufour E., Forestini M. 2000, A\&A, 358, 593 
\filbreak


\bibl
Silverstone, M.D., 2000, PhD thesis, UCLA
\filbreak

\bibl 
Skrutskie, M.F.,  Cutri, R.M. et al., 2006, AJ, 131, 1163
\filbreak

\bibl
Smith B.A. \& Terrile, R. J., 1984, Science, 226,1421-1424.
\filbreak

\bibl
Smith, P.S., Hines, D.C., Low, F.J.,  Gehrz, R.D., Polomski, E.F., and Woodward, C.E., 2006, ApJ, 644, L125
\filbreak

\bibl
Smith, R.,  Wyatt, M.C., \& Dent, W.R.F.,  2008, A\&A, 485, 897-915
\filbreak

\bibl
Spangler, C., Sargent, A.I., Silverstone, M.D., Becklin, E.E., Zuckerman, B., 2001, ApJ, 555, 932-944
\filbreak

\bibl
Su, K.Y.L.,  Rieke, G. H.,  Stapelfeldt, K.R., Stansberry, J.A.,  Bryden, G., Stapelfeldt, K.R., et al.  2006, ApJ, 653, 675
\filbreak

\bibl
Su, K.Y.L.,  Rieke, G. H.,  Stapelfeldt, K.R., Smith, P.S.,  Bryden, G., Chen, C.H., Trilling, D. E., 2008, ApJ, 679, 125
\filbreak

\bibl
Tacconi, L.J.,  et al., 2008, ApJ, 680, 246 
\filbreak

\bibl
Trilling, D. E., Bryden, G., Beichman, C. A., Rieke, G. H., Su, K. Y. L., Stansberry, J. A., et al., 2008, ApJ, 674, 1086 
\filbreak

\bibl
Tsiganis, K., Gomes, R.,  Morbidelli, A., Levison, H.F., 2005, Nature, 435, 459.
\filbreak

\bibl
Vorobyov, E.I., \&  Basu, S., 2008, ApJ, 676, L139-L142
\filbreak

\bibl
Voss, H., Bertoldi, F., Carilli, C., Owen, F.N., Lutz, D., Holdaway, M., Ledlow, M., and  Menten, K.M., 2006, A\&A, 448, 823
\filbreak

\bibl
Wilner, D. J.,Holman, M. J., Kuchner, M. J., Ho, P. T. P., 2002, ApJ,  569, L115
\filbreak

\bibl
Wyatt, M.C., Dent, W.F.R.,  2002, MNRAS, 334, 589.
\filbreak

\bibl
Wyatt, M.C., Dent, W.F.R.,  \& Greaves, J.S.,  2003, MNRAS, 342, 876.
\filbreak

\bibl
Wyatt, M.C., 2003, ApJ, 598, 1321
\filbreak

\bibl
Wyatt, M.C., 2006, ApJ, 639, 1153
\filbreak

\bibl
Wyatt, M.C., 2008, ARA\&A, 46, 339-383
\filbreak

\bibl
Zuckerman, B., 2001, ARA\&A, 39, 549
\filbreak

\bibl
Zuckerman, B., \& Inseok Song,  2004a, ARA\&A, 42, 685
\filbreak

\bibl
Zuckerman, B.,  \& Inseok Song, 2004b, ApJ, 613, L65
\filbreak

\end{document}